\documentclass[%
 reprint,
 amsmath,amssymb,
 aps,
]{revtex4-1}

\usepackage[]{graphicx}% Include figure files
\usepackage{dcolumn}% Align table columns on decimal point
\usepackage{bm}% bold math
\usepackage{amsmath}
\usepackage{amsthm} %for dealing with theorem environments,
\usepackage{mathtools}
\usepackage{algorithm}
\usepackage[noend]{algpseudocode}
\usepackage{color}
\usepackage[dvipsnames]{xcolor}
\usepackage{tikz}
\usepackage{enumitem}
\usepackage{mathtools}
\usepackage{subfigure}
\usepackage{placeins}

\newcommand{\avg}[1]{\left\langle #1 \right\rangle}
\newcommand{\bigAvg}[1]{\big\langle #1 \big\rangle}
\newcommand{\avgk}{\avg{k}}
\newcommand{\secMomK}{\avg{k^2}}
\newcommand{\cavgk}{\bar{k}}
\newcommand{\maxk}{k_{\mathrm{max}}}
\newcommand{\mink}{k_{\mathrm{min}}}
\newcommand{\maxkB}{k^b_{\mathrm{max}}}
\newcommand{\partialt}{\frac{\partial}{\partial t}}
\newcommand{\vecm}{\bold{m}}
\newcommand{\Vecm}{\bold{M}}
\newcommand{\betaS}{\beta^{ss' \rightarrow ss''}}
\newcommand{\betaSn}{\beta^{ss' \rightarrow ss_n}}
\newcommand{\betaSnn}{\beta^{ss_n \rightarrow ss'}}
\newcommand{\indSumMinus}{\sum_{R_{I,i} \in R^{s^{-}}_{I}}}
\newcommand{\indSumPlus}{\sum_{R_{I,i} \in R^{s^{+}}_{I}}}
\newcommand{\contSumMinus}{\sum_{R_{C,j} \in R^{s^{-}}_{C}}}
\newcommand{\contSumPlus}{\sum_{R_{C,j} \in R^{s^{+}}_{C}}}

\newcommand{\vecmS}{\vecm^{\{s'^+, s''^-\}}}
\newcommand{\vecmSn}{\vecm^{\{s_n^+, s''^-\}}}

\newcommand{\sumOverM}{\sum_{\vecm \in \Vecm_k}}

\newcommand{\rjonek}{{r_{j_1}, k}}
\newcommand{\rjtwo}{r_{j_2}}

\newcommand{\firstMoment}[2]{{\mu}_{#1} ( #2 )}
\newcommand{\secMoment}[3]{{\mu}_{#1} (#2, #3 )}
\newcommand{\secCentralMoment}[3]{\bar{\mu}_{#1} (#2, #3)}

\newcommand{\sumB}{\sum_{b\, \in B}}
\newcommand{\sumKBin}{\sum_{k \in b}}
\newcommand{\sumMk}{\sum_{\vecm \in \Vecm_k}}

\begin{document}

\preprint{APS/123-QED}

\title{Lumping of Degree-Based Mean-Field and Pair Approximation Equations \\ for Multi-State Contact Processes}% Force line breaks with \\

\author{Charalampos Kyriakopoulos$^1$}
\author{Gerrit Grossmann$^1$}
\author{Verena Wolf$^1$}
\author{Luca Bortolussi$^2$}
\affiliation{$^1$Computer Science Department, Saarland University}
\affiliation{$^2$Department of Mathematics and Geosciences,
 University of Trieste
}

\date{\today}% It is always \today, today,

\begin{abstract}
Contact processes form a large and highly interesting class of dynamic processes on networks, including epidemic and information spreading networks. While {devising} stochastic models of such processes is relatively easy, analyzing them is very challenging from a computational point of view, particularly for large networks appearing in real applications. One strategy to reduce the complexity of {their} analysis is to rely on approximations, often in terms of a set of differential equations capturing the evolution of a random node, distinguishing nodes with different topological contexts (i.e., different degrees of different neighborhoods), such as degree-based mean-field (DBMF), approximate master equation (AME), or pair approximation (PA) approaches.  The number of differential equations so obtained is typically proportional to the maximum degree $\maxk$ of the network, which is much smaller than the size of the  master equation of the underlying stochastic model, yet numerically solving these equations can still be problematic for large $\maxk$.
 In this paper, we consider AME and PA, extended to cope with multiple local states, and we provide an aggregation procedure that clusters together nodes
having similar degrees,  treating those in the same cluster as indistinguishable, thus reducing the number of equations while preserving an accurate description of global observables of interest.  We also provide an automatic way to build such equations and to identify a small number of degree clusters that give accurate results. 
The method is tested on several case studies, where it shows a high level of compression and a reduction of computational time of several orders of magnitude for large networks, with minimal loss in accuracy.

%\begin{description}
%\end{description}
\end{abstract}

\pacs{Valid PACS appear here}% PACS, the Physics and Astronomy
\maketitle

%\tableofcontents

\section{\label{sec:level1}Introduction}

Dynamic processes  on complex networks have attracted a lot of interest in recent years \cite{barabasi2002,newman2003,strogatz2001,palla2005,boccaletti2006,barrat2008,dorogovtsev2008,satoras2015}. Their behavior is complex and 
their dynamics is strongly influenced by the network topology through interactions between connected  nodes. Even more challenging is the study of adaptive networks, where both the node state and the network topology coevolve in a feedback relationship \cite{gross2006, gross2008, marceau2010, van2010, wu2012, wang2015}.

Among dynamic processes happening on networks, a prominent role is played by contact processes, which include information spreading and epidemic models \cite{barrat2008, mata2014, gomez2011, juhasz2012, boguna2009,  lee2013, satoras2015}.
In a contact process, each node of the network can be in one of several states, and neighboring nodes can interact and change their state. Interactions typically happen at random times, governed by an exponential distribution.
The underlying stochastic process is therefore a discrete-state Markov jump process. 

The analysis of such models is  very challenging, because of the huge state space of the underlying process: 
each possible network configuration (i.e.,~each assignment of states to nodes) represents a different discrete (global) state of the process. The exponential blowup of the state space rules out the use of any numerical technique to compute probabilities, such as uniformization or finite projection methods \cite{didier2009fast,wolf2010solving}, but also renders  simulation quite expensive, particularly for very large networks. An alternative and viable approach for large networks is provided by 
mean-field-like approximations \cite{bortolussi2013}. There are several forms and variations of such approximations \cite{satoras2001,gleeson2011,cator2012,mata2013, demirel2014, satoras2015} for network models, all sharing the same basic idea, namely to treat all nodes with a similar local structure as indistinguishable, and write a differential equation for each class of nodes, modeling the evolution of a random node in time. Depending on how much information about the local structure is taken into account, we obtain more refined approximations, at the price of more equations to be solved. In the simplest case, we treat all nodes with the same local state as indistinguishable (mean-field (MF) approximation \cite{bortolussi2013,barrat2008}). Improvements can be obtained by grouping nodes with respect to their state and degree, called degree-based (described also as heterogeneous sometimes) mean-field (DBMF) \cite{satoras2001, castellano2010, gleeson2012}, while an even better approximation is achieved by also representing explicitly also the state of neighboring nodes, as introduced by Gleeson's approximate master equation (AME) for the binary state case \cite{gleeson2011}. By implicitly relying on a binomial assumption, these equations can be simplified to
yield the pair approximation (PA) \cite{gleeson2011}. 
Although for many network topologies AME and PA give significantly more accurate results \cite{gleeson2013}, solving the AME and PA, even for the binary state case considered by Gleeson, is much more expensive than DBMF.

For the general multi-state case that we consider here, the solution of  AME and PA  is usually infeasible
for many real networks, as we show that
 the number of equations  is  $\binom{\maxk + |S|}{|S|-1} (\maxk + 1)$ for AME
 and $(|S|^2+|S|)\maxk$
 for PA, where $\maxk$ is the maximum degree of the network and $|S|$ is the number of possible states of a node.

Typically, when analyzing dynamic processes on complex networks, the interest is in emergent behaviors that manifest at a  collective level. Hence,  aggregated quantities like the fraction of nodes in a certain state are typically monitored. However, the network topology, even if marginalized in such observables,  highly influences   their dynamic properties. 
Often, the necessary granularity of the analysis is on the level of DBMF or PA.
Here, we show that the level of detail of these approaches can be further decreased. 
Our lumping approach
shows that for an accurate approximation of 
the mean behavior of the nodes, the exact degree and state of the neighborhood need not be monitored. 
Instead, aggregated information is sufficient, which strongly lowers the complexity of the analysis.
We decrease the number of equations by several orders of magnitude without noticable loss in precision compared to 
PA or DBMF. Moreover, we are able to set up an automatic procedure to construct such compressed equations, which avoids altogether the solution of the full system of equations. Computational gains are impressive, up to five orders of magnitude for large networks (large $\maxk$). This increased performance can be extremely useful when equations have to be solved many times, for instance for parameter sweeping or for parameter identification tasks.

%\todo{refer to some real world networks of barrat et al pg 24 to say that $k_{max}$ can be really large and then especially the PA equations  and AME (although we dont aggregate this) can even be computationally infeasible --
%Maybe something like this:
%Gleeson’s novel approach although very elegant comes with the drawback that capturing the dynamics in more complex networks requires the numerical solution of a large set of coupled equations, which hinders the analysis of large network sizes. }

In this paper, we consider  AME and PA for the case of $|S|$ local states, providing explicitly the AME and the  derivation of PA in a similar fashion as in \cite{Fennell2015}. For the special case of contact processes, we arrive at an equation that involves the first and second moments of the multinomial distribution, which renders the equations computationally more tractable compared to~\cite{Fennell2015}.

Our major contributions are the following: 
 (i) we derive a lumped set of equations for the multi-state DBMF and  PA equations and propose an algorithm to effectively construct this set; and (ii) we provide a numerical study on several non-trivial examples, showing that our approach performs well for several classes of degree distributions, that the reduction becomes more and more effective as the maximum network degree increases, and  that non-linear effects in the dynamics are captured well by the reduced equations.
Furthermore, (iii) we produced and made publicly available an open-source Python implementation of our method, taking a compact description of the model as input and returning Python files containing the lumped (and the full) DBMF and PA equations as output, together with procedures to integrate them in time and to plot the results.

The paper is further organized as follows:
  In Section \ref{sec:ApproxEq}, we introduce contact processes and several approximations, including a generalization of AME and PA for $|S|$ local states. We derive lumped equations, given a clustering of degrees of the network, in Section \ref{sec:lumping}.
In Section \ref{sec:results} we propose a clustering strategy that yields the lumped equations resulting in 
  an accurate solution, and present several case studies. Finally, in Section \ref{sec:conclusions} we draw final conclusions.

%\subsubsection{Wide text (A level-3 head)}
%The \texttt{widetext} environment will make the text the width of the
%full page, as on page~\pageref{eq:wideeq}. (Note the use the
%\verb+\pageref{#1}+ command to refer to the page number.) 
%\paragraph{Note (Fourth-level head is run in)}
%The width-changing commands only take effect in two-column formatting. 
%There is no effect if text is in a single column.

\section{Approximate Equations} \label{sec:ApproxEq}

In this section, we first introduce contact processes on networks, and then discuss several mean-field approaches to approximate their dynamics. These encompass mean-field, degree-based mean-field, approximate master equation, and pair approximation. The latter two methods are generalized from \cite{gleeson2011} to models with $|S|$ local states. 

%A similar generalization was developed independently in the PhD dissertation of Fennell \cite{Fennell2015}.

\subsection{Contact processes on networks}
A contact process among individuals on a finite graph $G$ can be described as an interaction model $M = \{S, R_I, R_C, G\},$ where $S = \{s_1, s_2, \ldots, s_{|S|}\}$ contains the possible states of an individual, $R_I = \{R_{I,1},  \ldots, R_{I, N_I}\}$ is the set of independent rules and $R_C = \{R_{C,1},  \ldots, R_{C,N_C}\}$ is the set of the contact rules among individuals. 

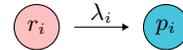
\begin{figure}
	\vspace{0.5cm}
	\begin{center}
		\begin{tikzpicture}[scale=0.1]
		\tikzstyle{every node}+=[inner sep=0pt]
		\draw [black, fill = pink] (22.2,-23.1) circle (3);
		\draw (22.2,-23.1) node {$r_i$};
		\draw [black, fill = SkyBlue] (39.4,-23.1) circle (3);
		\draw (39.4,-23.1) node {$p_i$};
		\draw [black] (27.2,-23.1) -- (34.4,-23.1);
		\fill [black] (34.4,-23.1) -- (33.6,-22.6) -- (33.6,-23.6);
		\draw (30.8,-22.6) node [above] {$\lambda_i$};
		\end{tikzpicture}
	\end{center}
	\caption{Independent rule $R_{I,i}$ changes a node's state $r_i$ to $p_i.$
	\label{fig:indeprule}  
}
\end{figure}

An independent rule $R_{I,i} \in R_I, \;i \in \{1, \ldots, N_I \}$ models a spontaneous state change of a 
  node of  the graph  $G$  from $r_i$ to $p_i$, where both $r_i, p_i \in S $ and $r_i \neq p_i$. 
  One instance of such a state change happens at an average rate of $\lambda_i$ per node.
Independent rules are illustrated in Figure \ref{fig:indeprule}.

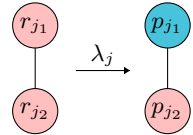
\begin{figure}
	\begin{center}
		\begin{tikzpicture}[scale=0.1]
		\tikzstyle{every node}+=[inner sep=0pt]
		\draw [black, fill = pink] (21.8,-23.1) circle (3);
		\draw (21.8,-23.1) node {$r_{j_1}$};
		\draw [black, fill = SkyBlue] (39.4,-23.1) circle (3);
		\draw (39.4,-23.1) node {$p_{j_1}$};
		\draw [black, fill = pink] (21.8,-34.5) circle (3);
		\draw (21.8,-34.5) node {$r_{j_2}$};
		\draw [black, fill = pink] (39.4,-34.5) circle (3);
		\draw (39.4,-34.5) node {$p_{j_2}$};
		\draw [black] (21.8,-26.1) -- (21.8,-31.5);
		\draw [black] (39.4,-26.1) -- (39.4,-31.5);
		
		\draw [black] (27.2,-29.1) -- (34.4,-29.1);
		\fill [black] (34.4,-29.1) -- (33.6,-28.6) -- (33.6,-29.6);
		\draw (30.8,-28.6) node [above] {$\lambda_j$};
		\end{tikzpicture}
	\end{center}
	\caption{Contact rule $R_{C,j}$ changes the state $r_{j_1}$ of a node to state $p_{j_1}$, whereas the state $r_{j_2} = p_{j_2}$ of the other node stays the same.}
	\label{fig:contactrule}
\end{figure}

A contact rule $R_{C, j} \in R_{C}, \;j \in \{1, \ldots, N_C \}$ describes the interaction of two neighboring nodes, respectively in states  $r_{j_1}, r_{j_2} \in S$, and changes the state of one such node, say state   $r_{j_1}$,  to   $p_{j_1} \in S$, while the state of the other node remains the same, i.e, $  r_{j_2}  =  p_{j_2}$. 
The general form of a contact rule is depicted in Figure \ref{fig:contactrule}. 
By abuse of notation, we define the average rate of this event per interacting edge as $\lambda_j$, i.e., 
we omit subscripts $I$ and $C$ to distinguish rates of  independent and contact rules.   
%
%
%On the other hand, the activation of a contact rule $R_{C, j} \in R_{C}, \;j \in \{1, \ldots, N_C \}$ with rule rate constant $\lambda_j$ demands the interaction of two neighboring nodes with states $r_{j_1}, r_{j_2} \in S$. 
%The result of this interaction is that w.l.o.g. node $r_{j_1}$ changes its state to $p_{j_1} \in S,$ while $r_{j_2}$ node retains its state, i.e, $  r_{j_2}  =  p_{j_2}.$  \ref{fig:contactrule}

%In order to count later properly the multiplicities of symmetric contact rules, i.e. those which either produce or consume two states of the same kind, it is convenient to distinguish between the upper, $R_{C,j_1}: \;r_{j_1} \rightarrow p_{j_1}$, and the lower part of the rule $R_{C,j_2}: \;r_{j_2} \rightarrow p_{j_2}.$

To express different variants of approximate master equations for an individual in state $s \in S$,
we   consider the subset of either the independent or the contact rules that affect the individual's  state. 
Let us define   $R^{s^{+}}_{I} = \{R_{I,i} \in R_I\mid  p_i = s \}$ as the subset of independent rules according to which a node spontaneously changes its state to $s.$ 
Similarly, $R^{s^{-}}_{I} = \{R_{I,i} \in R_I \mid  r_i = s  \}$ is the subset of independent rules where a node in  state $s$ 
changes  to some other state.
For contact rules that affect the population of individuals in state $s$, we define $R^{s^{+}}_{C} = \{R_{C,j} \in R_C\mid p_{j_1} = s\}$ for those contact rules that add a new individual in state $s$ and $R^{s^{-}}_{C} = \{R_{C,j}\in R_C \mid r_{j_1} = s  \}$ for those rules that result in a loss of an $s$ node.
%For contact rules that affect the population of individuals in state $s$ we define
%\vw{in the def of the set: I would add an $\exists \ell$} $R^{s^{+}}_{C} = \{R_{C,j} \mid (r_{j_\ell} \neq s  \wedge p_{j_\ell} = s)\}$ for those contact rules that produce a new individual in state $s$ and $R^{s^{-}}_{C} = \{R_{C,j_\ell} \mid (r_{j_\ell} = s  \wedge p_{j_\ell} \neq s)\}$ for those parts that result in a loss of an $s$ node.

The above contact process can be interpreted as  a continuous-time Markov chain 
$\{X(t),t\ge 0\}$ with state space $L$, where each global state $g \in L$ corresponds to  a labeled version of the graph 
$G$. 
The nonzero transition rates between the different  global states are given by the rate of either an independent or a contact rule, where we assume that a global state change is triggered by at most one rule. 
%\vw{the following argumentation/justification should go into the intro}
Since   the size of   $L$ is   $|S|^{N},$ where $N$ is the number
of nodes in $ G$, the analysis of $X$ is notoriously difficult. 
Typically, Monte Carlo simulations are carried out or
 equations for the approximation of its mean behavior are derived, since standard numerical approaches are infeasible.

\subsection{Mean-Field Equations}
One of the simplest assumptions to derive approximations of $X$ is that the nodes of the network are independent and equivalent. 
This is known as the mean-field assumption and gives rise to the well-known mean-field (MF) equation \cite{barrat2008}.
%%% Numbered equation

\begin{eqnarray}
\label{mf}
\frac{\partial x_s}{\partial t} &=&\sum_{R_{I,i} \in R^{s^{+}}_{I}} 
\lambda_i  x_{r_i} -\sum_{R_{I,i} \in R^{s^{-}}_{I}} \lambda_i  x_s\nonumber \\
 &+& \sum_{R_{C,j} \in R^{s^{+}}_{C}} \lambda_j  x_{r_{j_1}} x_{r_{j_{2}}}\langle k \rangle\ \\
 &-&\sum_{R_{C,j} \in R^{s^{-}}_{C}} \lambda_j  x_s x_{r_{j_{2}}} \avgk, \;\; \forall s\in S.\nonumber
\end{eqnarray}
In~\eqref{mf}, $x_s(t)$ is the fraction of nodes in state $s$ at time $t$,
$\avgk$ is the average degree of the network and $\lambda_i, \lambda_j$ are the rates associated with the rules $R_{I,i}, R_{C, j}$, respectively. 
The first two sums express the cases where a node changes its state to or from $s$ due to an independent rule and the last two sums 
represent every contact rule $R_{C,j}$  that either produces or consumes an $s$ node. The MF 
assumption is known to approximate well the dynamics of contact processes on either constant or 
Erdos-Renyi random graphs, where the degree of each node does not deviate much from the average degree \cite{barrat2008}.

If the network topology is slightly more complicated, which is the case for most real-world networks, then the MF
 equations cannot capture the dynamics of the spreading process anymore. 
In such networks, the infection probability of a node  depends on the node's degree.
Consequently, the approximation has to take into account the network's degree distribution $P$. 
This is particularly important for power law degree distributions, which are known to have a profound impact 
on the dynamics  of contact processes \cite{barrat2008, brodera2000, newman2001,clauset2009,satoras2015}. 
An approximation  that distinguishes between the different node degrees in the network, while 
averaging over all nodes   
that have the same degree and state, is the degree-based mean-field (DBMF) \cite{satoras2001}.
Let $K = \{1, \ldots, k_{\rm{max}} \}$  be the set of node degrees of the network and let $x_{s,k}$ be the fraction of nodes of degree $k$ that are in state $s \in S.$ 
We assume that the network has no nodes of degree zero, as they are irrelevant for the 
contact dynamics.
Then, for general contact processes, we have 
for all  $s\in S$ and all $k \in K$ that
\begin{eqnarray}
\label{dbmf}
\frac{\partial x_{s,k}}{\partial t}  &=&\sum_{R_{I,i} \in R^{s^{+}}_{I}} \lambda_i  x_{r_i, k} -\sum_{R_{I,i} \in R^{s^{-}}_{I}} \lambda_i  x_{s,k} \nonumber \\
&+& \sum_{R_{C,j} \in R^{s^{+}}_{C}} \lambda_j  x_{r_{j_{1}}, k} k p_k[r_{j_{2}}] \\ &-&\sum_{R_{C,j} \in R^{s^{-}}_{C}} \lambda_j  x_{s, k} k p_k[r_{j_{2}}]. \nonumber
\end{eqnarray}
Here, $p_k[s'] = Pr(s' \mid k)$  denotes the probability that a random neighbor of a degree $k$ node is in state $s' \in S$.
 We also let $p_k$ be the vector with entries $p_k[s']$, $ s'\in S$. If the network is uncorrelated, this probability is by definition independent of   
$k$ and for all $k \in K$  it holds (see also Appendix \ref{uncorellated}):  
\begin{equation}
\label{eq:probs}
p_k[s'] = p[s'] = \sum_{k' \in K} \frac{P(k') k' x_{s', k'}}{\avgk},
\end{equation}
where ${\avgk}$ is the average degree.
Hence, for uncorrelated networks the system of ODEs in~\eqref{dbmf} can be solved for some 
given degree distribution $P$ and initial conditions.
Having computed $x_{s,k}$ for all $k \in K,$ we can determine   $x_s = \sum_{k \in K} x_{s,k} P(k),$ which is usually  
  the fraction of interest.  

\subsection{Approximate Master Equation and Pair Approximation}\label{sec:AMEPA}
Mean-field theories (MF and DBMF) are known to be accurate for well-connected networks. 
However, they often  perform poorly on sparse networks \cite{gleeson2011}. 
A more accurate approximation is the approximate master equation (AME), which has been proposed for  binary state contact processes, such as the  SIS model of epidemic spreading \cite{gleeson2011,marceau2010}. 
Here we generalize the AME to contact processes with an arbitrary state space $S$.

For each $(s,k)$ node, i.e., each node that is in state $s$ and has degree $k,$  we   define a \emph{neighbor vector} $\vecm$  such that 
the  entry $\vecm[s_n]: 0 \leq \vecm[s_n] \leq k$ contains the number of edges that connect  the $(s,k)$ 
node   to a node in state $s_n$, that is, all entries
are non-negative integers and the sum of all entries is $k$.  Let $\Vecm_k$ be the set of all such vectors. 
For $x_{s,k,\vecm},$ i.e., the fraction of nodes of degree $k$ which are in state $s$ and have neighbor vector $\vecm$,  we get
for all $s\in S$, $k\in K$, and $ \vecm\in \Vecm_k,$

\begin{eqnarray}\label{ame}
&&\frac{\partial x_{s, k, \vecm}}{\partial t} = \sum_{R_{I,i} \in R^{s^{+}}_{I}} \lambda_i  x_{r_i, k,\vecm} -\sum_{R_{I,i} \in R^{s^{-}}_{I}} \lambda_i  x_{s, k, \vecm} \nonumber \\
&+& \sum_{R_{C,j} \in R^{s^{+}}_{C}} \lambda_j  x_{r_{j_1}, k, \vecm} \vecm[r_{j_2}] -\sum_{R_{C,j} \in R^{s^{-}}_{C}} \lambda_j  x_{s, k, \vecm}  \vecm[r_{j_2}] \nonumber \\ 
&+&\sum_{\substack{ (s',s'') \in S^2\\ s'\neq s''}}  \betaS x_{s, k,\vecmS} \vecmS[s'] \nonumber\\
&-& \sum_{\substack{ (s',s'') \in S^2\\ s'\neq s''}} \betaS x_{s, k,\vecm}  \vecm[s']. 
\end{eqnarray}
The first four terms in the above equation are similar to
those in Eq.\,\eqref{dbmf} and   express the cases where we either add or delete
 an $(s,k,\vecm)$ node via an independent or a contact rule. The difference now is that we can write down the precise number of edges between two reacting neighboring nodes in states $r_{j_1}, r_{j_2}$ instead of using their average number
  as in \eqref{dbmf}.

Each term in the last two sums represents the case where one of the $k$ neighbors of an $(s,k)$ node changes 
its state from $s'$ to $s''$. 
This change results in either obtaining a new $(s,k,\vecm)$ node (first sum) or losing such a node (second sum).

The vector $\vecmS$ is such that all entries are  equal to those of $\vecm$, apart from the $s'$-th entry, which is equal to $\vecm[s'] +1$, and the $s''$-th entry, which is equal to $\vecm[s''] -1.$ 
In addition, $\betaS$ is the average rate at which $ss'$ edges change to $ss''$ 
edges, where $s,s',s'' \in S.$ 
To derive an expression for $\betaS$ we  define the sets  $R^{s'\rightarrow s''}_{I} = \{R_{I,i} \in R_I\mid r_i = s'  \wedge p_i = s'' \}  $ as the set of independent rules and $ R^{s'\rightarrow s''}_{C} = \{R_{C,j} \in R_C \mid r_{j_1} = s'  \wedge p_{j_1} = s'' \}$  as the set of 
 contact rules that result in a node change from $s'$ to $s''$. 
Now, we   define the rate $\betaS$, which is independent of $k$, as 
\begin{eqnarray}
\label{betaS}
&&\betaS = \frac{1}{\avg{\sumMk \vecm[s] x_{s',k,\vecm}}}  \cdot \\
&&\avg{\sumMk \!\!\vecm[s] x_{s',k,\vecm} \Big(\sum_{R_{I,i}\in R^{s'\rightarrow s''}_{I} }\!\!\!\! \lambda_i  + \!\!\!\sum_{R_{C,j}\in R^{s'\rightarrow s''}_{C}} \!\!\!\!\!\lambda_j \vecm[r_{j_2}] \Big) }, \nonumber 
\end{eqnarray} 
where the above average is taken over all degrees of the network, i.e., $\avg{f(k)} =  \sum_{k=1}^{\maxk} P(k) f(k).$ 
%
% Doestroys the layout somehow
%\gerrit{$M_k = \{ m \in \mathbb{Z}_{\geq 0}^{|S|} \mid \sum_{i=0}^{|S|} m_i = k \}$}
%

%In order to calculate $\betaS$ we count the number of $ss'$ edges in the network at time $t$ and the number of $ss'$ edges that change to $ss''$ edges within time interval $dt$ and we get the ratio of the latter to the former, i.e., $$\betaS = \frac{ \#(ss' \rightarrow ss'') \mathrm{within} \;\partial t } {\# ss'}.$$  It holds that $\# ss' = \langle \sum_{\vecm \in M} \skm \vecm[s'] \rangle,$ where $M = \{ \vecm: \sum_{i=1}^{n} \vecm[i] = k \}$ and \\ $\#(ss' \rightarrow ss'') = \Big\langle \sum_{\vecm \in M} \Big [ \sum_{R_{I,i} \in \RiS} \skm m[s'] \lambda_i + \sum_{R_{C,j_\ell} \in \RcS} \skm \vecm[s'] \lambda_j m[j_{3-\ell}] \Big ] \Big\rangle, $ where $\langle y(k) \rangle$ for some function $y$ stands for $\sum_{k=0}^{\maxk} = P(k) y(k).$

The number of differential equations that need to be solved for the   general case of the AME is $|S| \sum_{k=0}^{\maxk}  \binom{k+ |S| -1}{|S|-1}  = \binom{\maxk + |S|}{|S|-1} (\maxk + 1).$ The binomial sum arises from the number of ways that, for a fixed degree $k,$ one can distribute $k$ unlabeled edges to $|S|$ states (see also Appendix \ref{numberOfAME}).
For the special case of $|S|=2,$ the number of equations is $(\maxk+1) (\maxk + 2)$   \cite{gleeson2011}.  

A coarser approximation, the pair approximation \cite{gleeson2011}, is possible if instead of considering one equation for each vector  $\vecm$ we assume a multinomial distribution for the number of edges that a node  of degree $k$ in state $s$ has to all other states. This way, one needs to consider a (time-dependent) vector $p_{s, k}$ whose $i-$th entry    $p_{s, k}[s_i]$ is defined as the probability that a random neighbor of an $(s,k)$ node is in state $s_i \in S$. Here we consider the general pair approximation  for contact processes with $|S|$ states, generalizing  the binary state approach of \cite{gleeson2011} similarly to \cite{Fennell2015}. A detailed derivation is provided in  Appendix \ref{AMEToPAProof}. 
Assuming that $x_{s,k}>0$, we have for all 
$s\in S$, $k \in K $, $s_n \in S$,
%f\big(\vecm; p(s_k)\big)
\begin{eqnarray}
\label{PApart1}
&&\frac{\partial x_{s, k}}{\partial t}   =  \sum_{R_{I,i} \in R^{s^{+}}_{I}} \lambda_i   x_{r_i, k} -\sum_{R_{I,i} \in R^{s^{-}}_{I}} \lambda_i  x_{s, k} \\ 
&&+ \sum_{R_{C,j} \in R^{s^{+}}_{C}} \lambda_j x_{r_{j_1},k}\firstMoment{r_{j_1},k }{r_{j_2}} -\sum_{R_{C,j} \in R^{s^{-}}_{C}} \lambda_j x_{s,k} \firstMoment{s,k}{r_{j_2}} \nonumber 
\end{eqnarray}
and
\begin{eqnarray}
\label{PApart2}
&&
\frac{\partial p_{s,k}[s_n] }{\partial t}=  - \frac{\partial x_{s, k}}{\partial t}  \frac{p_{s,k}[s_n]}{x_{s,k}} \nonumber \\
&&+ \indSumPlus \frac{\lambda_i x_{r_i,k}}{x_{s,k} k} \firstMoment{r_{i,k}}{s_n} - \indSumMinus \frac{\lambda_i}{ k } \firstMoment{s,k}{s_n} \nonumber\\
&&+ \contSumPlus \!\!\!\frac{\lambda_j x_\rjonek} {x_{s,k} k} \secMoment{\rjonek}{\rjtwo}{s_n}  - \!\!\!\!\!\contSumMinus \frac{\lambda_j }{ k} \secMoment{s,k} {\rjtwo }{s_n} \nonumber \\
&&+ \sum_{\substack{ s' \in S\\ s'\neq s_n}}  \betaSn p_{s,k}[s'] - \sum_{\substack{ s' \in S\\ s'\neq s_n}}  \betaSnn p_{s,k}[s_n],%\end{split}
\end{eqnarray}
where $\firstMoment{s, k}{s_n} =  \mathbb{E}\left[ \vecm[s_n] \right]$     %\sumMk f(\vecm, p_{s,k}) \vecm[s_n]$ 
is the expected number of $s_n$ neighbors of an $(s, k)$ node. We assume that the   number 
of $s_n$ neighbors is   binomially distributed, i.e., the distribution is $\mathrm{\textbf{B}}(k, p_{s,k}[s_n])$. 
Furthermore, $\secMoment{s,k}{s_n}{s'_n} =  \mathbb{E}\left[ \vecm[s_n]\vecm[s'_n] \right]$ is the second mixed moment of the 
corresponding multinomial distribution with parameter vector
% $\mathrm{\textbf{Mult}}(k, \big[ p_{s,k}[s_n], p_{s,k}[s'_n] \big] )$, 
  $p_{s, k}.$ 
  In a similar fashion, exploiting the multinomial assumption  
  we can further simplify (see Appendix \ref{AMEToPAProof})  $\betaS$ to 
 \small
\begin{eqnarray}
\label{betaS}
&&\betaS = 1/{\avg{x_{s',k} \firstMoment{s',k}{s}}} \cdot  \big(\bigAvg{x_{s',k} \sum_{R^{s'\rightarrow s''}_{I}} \lambda_i  \firstMoment{s',k}{s}} \nonumber\\
&& + \ \bigAvg{x_{s',k}  \sum_{R^{s'\rightarrow s''}_{C}} \lambda_j \secMoment{s',k}{s}{\rjtwo} }\big) . 
\end{eqnarray}
\normalsize
The number of equations is now reduced to $(|S|^2+|S|)\maxk$, where $|S|$ is typically much smaller than $\maxk.$
In addition, the use of the moments of the multinomial distribution results in an enormous computational gain compared
to the enumeration of huge sums over vectors when solving the above ODE system.  For a detailed derivation of the above equations and a proof of their correctness we refer the reader to Appendix \ref{AMEToPAProof}.

%\begin{remark}
As   in the pair approximation for the binary case \cite{gleeson2011}, 	the right hand side of Eq.\,\eqref{PApart2} is undefined if  $x_{s,k}=0$. 
However, it is easy to check that if $x_{s,k} \neq 0,$ then the system can never reach such a value, as the negative terms in Eq.\,\eqref{PApart1} depend linearly on  $x_{s,k}$. 
In the following, without loss of generality, we   assume that  $x_{s,k}\neq 0$ for all $s$ and $k$ at 
all times. An alternative is to impose that $\partialt p_{s,k}[s_n] = 0$ whenever $x_{s,k}=0$, since there are no nodes of degree $k$ in state $s$. 

\section{Lumping Equations} \label{sec:lumping}
In this section we present aggregated equations  for DBMF and PA.
We will first   explain all  assumptions that are necessary for the lumping, 
when we partition variables into bins obtained by aggregating degrees according to the degree distribution. 
We will discuss how to obtain a good partition of the degree set $K$ in Section \ref{sec:results}, together with experimental results on several case studies. 
An important issue of the proposed method is that our aim is to approximate global averages such as the total fraction of nodes that are in a certain state in a given time interval, and not more detailed values such as the probability of (a  set of) certain global states of the underlying Markov chain $X$.

\subsection{Lumping of Degree-Based Mean-Field Equations}
Our next goal is to derive a set of differential equations based  on an aggregation of the DBMF equations (see Eq.\,\eqref{dbmf}) in order to approximate  the total population fraction for each state $s$.
The starting point is a partition $B = \{b_1,\ldots,b_h\} \subset 2^K$  of the degrees $K=\{1,\ldots,\maxk\}$, i.e., such that the disjoint union    $\dot{\bigcup}_{b\in B} b $ yields $K$ and $\emptyset \not\in B$. A set of degrees $b\in B$ will be called a bin.
Let $x_{s,b}$ denote the corresponding fraction of nodes in state $s$ that have a degree in bin $b$ (also called $(s,b)$ nodes in the following). Knowing the fractions $x_{s,b}$ for all $b \in B,$ we can simply compute $x_s = \sum_{b \in B} x_{s,b} P(b)$ to get the total fraction of  nodes in state $s$. Here,    $P(b) =  \sumKBin P(k) $ is the degree probability of bin $b$. Note that if a network is finite and has $N$ nodes, we can always write $x_{s,b} = \frac{X_{s,b}}{N_b}, $ where $X_{s,b}$ is the number of  $(s,b)$ nodes  and $N_b$ is the total number of nodes that have degree 
$k\in b$, i.e., $P(b) =  N_b/N$.  We remark that $x_{s,b} = \sum_{k \in b} x_{s,k} P(k|b)$.

%We need, thus, to find a partition from the set of the original variables, which are the network degrees $k \in K,$ to the lumped ones, which are the bins $b \in B.$ 
%We call this partition binning and the corresponding lumped variables bins. 
In the next section,  we  will describe a heuristic to determine a suitable partition $B$. 
For now, we assume that $B$ is given and that $P(b) \neq 0$ for all $b \in B$. 
In this case, the probability $P(k|b)  =P(k)/P(b)$ that a random node that belongs to bin $b$ is of degree $k$ is well defined.
Multiplying both sides of Eq.\,\eqref{PApart1}  with   $P(k|b)$ and summing over all degrees $k \in b$,
   we get the following equation for $s\in S$ and $b \in B$.  
\begin{eqnarray}
\label{dbmfFraction}
&&\partialt \sumKBin x_{s,k}  P(k|b) = \nonumber\\
&&\sum_{R_{I,i} \in R^{s^{+}}_{I}} \lambda_i  \sumKBin x_{r_i, k}  P(k|b) -\sum_{R_{I,i} \in R^{s^{-}}_{I}} \lambda_i  \sumKBin x_{s,k}  P(k|b) \nonumber\\
&&+ \sum_{R_{C,j} \in R^{s^{+}}_{C}} \lambda_j p[r_{j_{2}}] \sumKBin x_{r_{j_{1}, k}} k  P(k|b) \nonumber\\
&&-\sum_{R_{C,j} \in R^{s^{-}}_{C}} \lambda_j  p[r_{j_{2}}] \sumKBin x_{s, k} k  P(k|b)
\end{eqnarray}
Next we replace sums of the form $\sum_{k \in b} x_{s,k} P(k|b)$   by the lumped fraction $x_{s,b}$. 
 However,   exact lumping is not possible for  terms of the form $\sum_{k \in b}  x_{s, k} k P(k|b)$. 
 In order to express the last two sums in terms of the lumped fractions, we 
 make the following assumption: 
\begin{equation}
	\label{assum}
\parbox{0.8\textwidth/ 2}{\it {\bf (Homogeneity)} For all $k\in b$, the fraction $x_{s,b}$ of degree $b$ nodes that are in state $s$
is  equal to the fraction $x_{s,k}$ of degree $k$ nodes that are in state $s$.} \end{equation}
In other words, we assume that in bin $b$  the fraction of nodes in state $s$ is the same for each degree $k \in b$. 
Now, \eqref{dbmfFraction} becomes
%In terms of populations this is equivalent to: $x_{s,k} = x_{s,b} \; \iff \frac{X_{s,k}}{N_k} = \frac{X_{s,b}} {N_b} \iff X_{s,k} = \frac{P(k)}{P(b)} X_{s,b}.$ This way we get:
\begin{eqnarray}
\label{dbmfFractionFinal}
  &&\partialt  x_{s,b}   = \sum_{R_{I,i} \in R^{s^{+}}_{I}} \lambda_i  x_{r_i, b}   -\sum_{R_{I,i} \in R^{s^{-}}_{I}} \lambda_i   x_{s,b}   \\
&+& \!\!\!\!\sum_{R_{C,j_\ell} \in R^{s^{+}}_{C}} \lambda_j p[r_{j_{2}}] x_{r_{j_{1}},b} \avgk_b  \!-\!\!\!\!\sum_{R_{C,j} \in R^{s^{-}}_{C}} \!\!\lambda_j  p[r_{j_{2 }}] x_{s, b} \avgk_b,  \nonumber 
\end{eqnarray}
where $\avgk_b = \sum_{k\in b} k P(k|b)$ is the average degree in bin $b$.
Since the probability $p[s_n]$ depends on the fractions $x_{s,k}$,
we must also express it in terms of   lumped fractions.
Imposing Assumption \eqref{assum} to \eqref{eq:probs} yields 
\begin{align}\label{dbmfProbability}
\begin{split}
p[s_n] &= \frac{1}{\avgk} \sum_{k' \in K} P(k') k' x_{s_n, k'} \\
& = \frac{1}{\avgk}  \sum_{b \in B} \sum_{k' \in b} P(k') k' x_{s_n, k'} \\
&= \frac{1}{\avgk } \sum_{b\in B} x_{s_n,b} P(b) \sum_{k' \in b}  P(k'|b)  k' \\
&= \frac{1}{\avgk } \sum_{b\in B} x_{s_n,b}  P(b) \avgk_b. 
\end{split}
\end{align}
Equations~\eqref{dbmfFractionFinal} and~\eqref{dbmfProbability} define the aggregated equations of the DBMF model. 
%\vw{Somewhere is a discussion necessary about (1) where the approximation is made
%and (2) the reduced number of equations and how that influences our
%choice of bins.}

\begin{emph}
Since Assumption \eqref{assum} is usually not true, we obtain   approximations of the true fractions $x_{s,b}$ and we use
them as approximations of $x_{s,k}$.  
%Obviously, the  approximation error depends on . Note that once we aggregate together the probability mass of binned variables, there is no way to know the original values. 
%Note that assumption \eqref{assum} can be justified by the principle of maximum entropy:  
%partitioning in equal amounts corresponds to maximizing the uncertainty in the reconstruction. 
\end{emph}

\subsection{Lumping of Pair Approximation Equations}
For the PA equations~\eqref{PApart1} and~\eqref{PApart2}, we also fix a partition $B$ of degrees into bins and 
  multiply  Eq.\,\eqref{PApart1} with   $P(k|b).$ Summing over all degrees $k \in b$, we get the following lumped equation for the fraction of nodes of bin $b$ that are in state $s.$
\begin{eqnarray}\label{PAFractionsLumped1}
&&\partialt \sumKBin x_{s,k} P(k|b) = \nonumber\\
 &&\, \sum_{R_{I,i} \in R^{s^{+}}_{I}} \lambda_i \sumKBin x_{r_i, k} P(k|b)  -\!\!\sum_{R_{I,i} \in R^{s^{-}}_{I}} \lambda_i  \sumKBin x_{s,k} P(k|b)  \nonumber\\
&&+ \sum_{R_{C,j} \in R^{s^{+}}_{C}} \lambda_j \sumKBin x_{r_{j_1}, k} \firstMoment{\rjonek}{\rjtwo} P(k|b) \nonumber \\
&& -\sum_{R_{C,j} \in R^{s^{-}}_{C}} \lambda_j \sumKBin x_{s, k} \firstMoment{s, k}{r_{j_2}}P(k|b), 
\end{eqnarray}
where $\firstMoment{\rjonek}{\rjtwo}$ is again the corresponding first moment of the 
multinomial distribution (cf.\! Section \ref{sec:AMEPA}).
%As before, we write $x_{s, b}$ for the fraction of $s$ nodes in bin $b$ and we define $p_{s,b}[s_n]$ as
 %the probability that a randomly selected neighbor of an $s$ node with degree $k\in b$ is in state $s_n \in S.$
Similarly to the DBMF lumping approach, it is necessary to make  simplifying assumptions
in order to write the last two sums of \eqref{PAFractionsLumped1} in terms of lumped fractions.
More specifically,  we again impose Assumption \eqref{assum} and in addition
we assume that the probability  that an $(s,k)$ node has an $s_n$ neighbor is equal for all $k\in b$, i.e.,  for all  $k\in b$,
$p_{s,b}[s_n]$ is defined as the probability that a randomly selected neighbor of an $(s,k)$ node  is in state $s_n \in S$. 
Hence, if $\firstMoment{s, b}{s_n} $ is the average number of $s_n$ neighbors of an $s$ node in bin $b$,
and if this number is binomially distributed, we get
\begin{eqnarray}
\label{eq:assumimpl}
\firstMoment{s, b}{s_n} &=&  \sumKBin \firstMoment{s, k}{s_n} P(k|b)=
\sumKBin k p_{s, k}{[s_n]} P(k|b) \nonumber \\
&=&p_{s, b}{[s_n]}\sumKBin k  P(k|b)=
p_{s, b}{[s_n]}\avgk_b.  
\end{eqnarray}
%\vw{Charis, you wrote that multinomial assumption is necessary here. Why?}  \todo{bec otherwise the first moment does not equal k *p}
%  With a small abuse of notation we can write $\firstMoment{s, b}{\vecm[s_n]} =  p_{s,b}[s_n] \avgk_b$ as the average number of $s_n$ neighbours of an $s$ node in bin $b$ (under the before mentioned assumptions).
%  
%  is equal for all $k \in b$ and thus can be written
%  as  $\firstMoment{s, b}{\vecm[s_n]}$ where $\firstMoment{s,b}{\vecm[s_n]}$ is the expected number of edges between an $s$ node in bin $b$ and a neighbor in state $s_n.$
%  
Consequently,~\eqref{PAFractionsLumped1} gives for all $s \in S$ and
$b\in B$,
\begin{eqnarray}
\label{PAFractionsLumped2}
\partialt x_{s,b} &=&  \sum_{R_{I,i} \in R^{s^{+}}_{I}} \lambda_i  x_{r_i, b} -\sum_{R_{I,i} \in R^{s^{-}}_{I}} \lambda_i  x_{s,b}   \nonumber\\
&+& \sum_{R_{C,j} \in R^{s^{+}}_{C}} \lambda_j  x_{r_{j_1}, b} \firstMoment{r_{j_1}, b}{\rjtwo}   \\
&-&\sum_{R_{C,j} \in R^{s^{-}}_{C}} \lambda_j  x_{s, b} \firstMoment{s, b}{r_{j_2}}. \nonumber
\end{eqnarray}
Finally,   using~\eqref{eq:assumimpl} we get
\begin{eqnarray}
\label{PAFractionsLumpedFinal}
 \partialt x_{s,b} &=& \sum_{R_{I,i} \in R^{s^{+}}_{I}} \!\!\lambda_i  x_{r_i, b}  -\sum_{R_{I,i} \in R^{s^{-}}_{I}} \!\!\lambda_i  x_{s,b}  \nonumber \\
&+& \sum_{R_{C,j} \in R^{s^{+}}_{C}} \!\!\!\lambda_j  x_{r_{j_1}, b} p_{r_{j_1}, b}[\rjtwo] \avgk_b  \\
&-& \sum_{R_{C,j} \in R^{s^{-}}_{C}} \!\!\!\lambda_j  x_{s, b} p_{s, b}[r_{j_2}] \avgk_b.\nonumber
\end{eqnarray}
%\begin{align}\label{PAFractionsLumpedFinal}
%\begin{split}
%\partialt x_{s,b} &=  \sum_{R_{I,i} \in R^{s^{+}}_{I}} \lambda_i  x_{r_i, b}    -\sum_{R_{I,i} \in R^{s^{-}}_{I}} \lambda_i  x_{s,b}   \\
%&+ \sum_{R_{C,j} \in R^{s^{+}}_{C}} \lambda_j  x_{r_{j_1}, b} \firstMoment{r_{j_1}, b}{\vecm[r_{j_2}]}    -\sum_{R_{C,j} \in R^{s^{-}}_{C}} \lambda_j  x_{s, b} \firstMoment{s, b}{\vecm[r_{j_2}]} .\\
%\end{split}
%\end{align}
To derive  a differential equation for the time-dependent variable $p_{s,b}[s_n]$ we multiply 
Equation \eqref{PApart2} with $P(k|b)$ and sum over all $k \in b$. Assuming $x_{s,k}>0$, this yields 
\begin{eqnarray}\label{PAProbLumped1}
&&\partialt \sumKBin p_{s,k}[s_n] P(k|b) = -\sumKBin\partialt x_{s,k} \frac{p_{s,k}[s_n]}{x_{s,k}} P(k|b) \nonumber\\
&+& \indSumPlus \lambda_i  \sumKBin \frac{x_{r_i,k}}{x_{s,k} } p_{r{_i, k}}[s_n] P(k|b) \nonumber \\
&-& \indSumMinus \lambda_i \sumKBin p_{s,k}[s_n] P(k|b) \nonumber\\
&+& \contSumPlus \lambda_j  \sumKBin \frac{x_\rjonek} {x_{s,k} k  } \secMoment{r_{j_1}, k}{\rjtwo}{s_n} P(k|b)   \nonumber\\ 
&-& \contSumMinus \lambda_j \sumKBin \frac{1}{k} \secMoment{s,k}{r_{j_2}}{s_n}  P(k|b) \nonumber\\
&+& \sum_{\substack{ s' \in S\\ s'\neq s_n}}  \betaSn \sumKBin p_{s,k}[s'] P(k|b) \nonumber \\
&-& \sum_{\substack{ s' \in S\\ s'\neq s_n}}  \betaSnn \sumKBin p_{s,k}[s_n] P(k|b). 
\end{eqnarray}     
Using \eqref{eq:assumimpl} as well as the above assumption
$p_{s,k}[s_n] = p_{s,b}[s_n]$ together with Assumption \eqref{assum}  we get: \small
\begin{eqnarray}
\label{PAProbLumpedFinal}
&&\partialt  p_{s,b}[s_n] = -\partialt x_{s,b} \frac{p_{s,b}[s_n]}{x_{s,b}}  + \indSumPlus \lambda_i  \frac{x_{r_i,b}}{x_{s,b} } p_{r{_i, b}}[s_n]  \nonumber \\
&-& \!\!\!\indSumMinus\!\!\! \lambda_i p_{s,b}[s_n] + \!\!\!\contSumPlus \!\!\!\!\lambda_j  \frac{x_{r_{j_1}, b}} {x_{s,b} } \sumKBin \frac{1}{k} \secMoment{r_{j_1},k}{r_{j_2}}{s_n}  P(k|b)  \nonumber \\ 
&-& \!\!\!\contSumMinus \!\!\!\lambda_j \sumKBin \frac{1}{k} \secMoment{s,k}{r_{j_2}}{s_n}  P(k|b)  \\
&+& \sum_{\substack{ s' \in S\\ s'\neq s_n}}  \betaSn  p_{s,b}[s']   - \sum_{\substack{ s' \in S\\ s'\neq s_n}}  \betaSnn  p_{s,b}[s_n]. \nonumber
\end{eqnarray}     \normalsize
Let $\avg{k-1}_b = \sum_{k\in b} (k-1) P(k|b) $. 
If $r_{j_2} \neq s_n,$ then (see also Appendix \ref{sumsInPALumping})
\begin{equation}\label{needed1}
  \sumKBin \frac{1}{k} \secMoment{s,k}{r_{j_2}}{s_n}  P(k|b)=\avg{k-1}_b p_{s,b}[r_{j_2}] p_{s,b}[s_n]
  \end{equation}
  and if $r_{j_2} = s_n$, then
\begin{equation}\label{needed2}
\sumKBin \frac{1}{k} \secMoment{s,k}{r_{j_2}}{s_n}  P(k|b)=  
 p_{s,b}[s_n]  + p_{s,b}[s_n]^2 \avg{k-1}_b.
   \end{equation}
 %   For the complete derivation of~\eqref{PAProbLumpedFinal} we refer again to the appendix. 
We remark that in Eq.\,\eqref{PAProbLumpedFinal}, the calculation of the average rates $\betaSn$ and $\betaSnn$ requires the 
detailed fractions $x_{s,k}$ (cf.\! Eq.\,\eqref{betaS}).
Hence, we define, similar to Eqs.\,\eqref{eq:assumimpl} and~\eqref{betaS},  the second moments   
\begin{equation}
\label{eq:assumimpl2}\secMoment{s,b}{s_n}{s_n'} =  \sumKBin \secMoment{s, k}{s_n}{s_n'} P(k|b),
\end{equation} 
and the  average rate
\small
\begin{eqnarray}
\label{betaLumped}
&&\betaS_{B} = 1/{\avg{x_{s',b} \firstMoment{s',b}{s}}_b}\\
&& \big({\bigAvg{x_{s',b} \sum_{R^{s'\rightarrow s''}_{I}} \lambda_i  \firstMoment{s',b}{s}}_b  + \bigAvg{x_{s',b}  \sum_{R^{s'\rightarrow s''}_{C}} \lambda_j \secMoment{s',b}{s}{\rjtwo} }_b}\big).  \nonumber
\end{eqnarray} \normalsize
Now, we can replace all unknown terms in \eqref{PAProbLumpedFinal} by using \eqref{needed1} and  \eqref{needed2},
as well as $\betaS=\betaS_{B}$, where the latter equality holds under Assumption \eqref{assum}, and
%Using~\eqref{eq:assumimpl2}  in $\betaS \;\forall (s',s'') \in S^2$  and after some calculations similar to the ones above, we get $\betaS_{\text{lumping}}$ as it is shown in~\eqref{betaLumped}.
then the lumped ODE system for the PA model is given by Eqs.\,\eqref{PAFractionsLumpedFinal} and \eqref{PAProbLumpedFinal}. 
%\small
%\begin{eqnarray}
%\label{betaLumped}
%&&\betaS_{\text{lumping}} = \\
%&&\frac{\bigAvg{x_{s',b} \sum_{R^{s'\rightarrow s''}_{I}} \lambda_i  \firstMoment{s',b}{s}}_b  + \bigAvg{x_{s',b}  \sum_{R^{s'\rightarrow s''}_{C}} \lambda_j \secMoment{s',b}{s}{\rjtwo} }_b} {\avg{x_{s',b} \firstMoment{s',b}{s}}_b} \nonumber
%\end{eqnarray} \normalsize
%\section{Figures \& Tables}

%The output for figure is:
%
%\begin{figure}[!h]
%%\centering\includegraphics[width=2.5in]{figurename.eps}
%%%%call your figure name in the place "figurename.eps"
%\caption{Insert figure caption here}
%\label{fig_sim}
%\end{figure}

%\begin{figure*}[!h]
%\centerline{\subfloat[Case I]\includegraphics[width=2.5in]{figurename.eps}%
%\hfil
%\subfloat[Case II]{\includegraphics[width=2.5in]{figurename.eps}%
%\label{fig_second_case}}}
%\caption{Simulation results}
%\label{fig_sim}
%\end{figure*}

\vskip2pc

\section{Binning Method and Results}
\label{sec:results}
In this section, we discuss strategies to choose the partition $B$, which strongly influences the quality of the approximation
of the lumped equations. We focus on an accurate approximation of the fraction of nodes in state $s$.
%We turn now to describe how to obtain a good clustering of degrees. Clearly, the best binning is not only model specific, but also %depends on the quantities we need to accurately preserve. In this paper, we focus on collective properties, mainly the fraction of %nodes in each state: $x_s, \forall s\in S$.

 Note that if we solve the original DBMF or PA equations we have $x_s(t) = \sum_{k\in K} x_{s,k}(t) P(k)$,
while from the lumped ODEs we get $\bar{x}_s(t) = \sum_{b\in B} \sumKBin x_{s,b}(t) P(k)$.
Hence, the error that we obtain at a certain time $t$ by solving the lumped ODEs instead of the full equations is 
$$
\begin{array}{rcl}
\epsilon_B(t) &=& || x_s(t) - \bar{x}_s(t) || \\[1ex]
 &=& ||\sumB \sumKBin x_{s,k}(t) P(k) -  x_{s,b}(t) P(k) || \\[1ex]
 &=& ||\sumB \sumKBin P(k) \underbrace{( x_{s,k}(t) - x_{s,b}(t) )}_{\epsilon_{k,b}(t)} ||.
\end{array}$$ 
 This leads to the   following   observations.
\begin{itemize}[leftmargin =*]
	\item If $P(k)$ is large, then the deviation $\epsilon_{k,b}(t)$  of $x_{s,b}(t)$ from $x_{s,k}(t)$ will have a great impact on the total error. On the other hand, if $P(k)$ is small, then one could handle even large deviations from the true solution.
	\item For a fixed bin,  the smaller the relative difference between the degrees (a more homogeneous    bin),
	the smaller is the error $\epsilon_{k,b}(t)$. For instance, we can expect that the fractions 
	$x_{s,500}, x_{s,501}, x_{s,502}$ behave much more similarly than the fractions 
	$x_{s,1}, x_{s,2}, x_{s,3}$ over time. 
	Intuitively, this comes from the fact that for
	nodes with only   one or two neighbors the chance of  
	triggering a contact rule is less similar than the chance of 
	nodes with 500 or 501 neighbors. 
	In other words, the relative difference of the time derivative for degrees $k$ and $k+1$ is smaller for larger degrees than for small ones. 
\end{itemize}
For a given partition $B$, we define the total error $\epsilon_{\text{tot},B}=\max_{t\le T_{max}}\epsilon_B(t),$ where $T_{max}$ is the maximum time horizon. 

\subsection{Hierarchical Clustering}

%The binning that we choose will of course influence $\epsilon_{k,b}, \forall k\in K$ and consequently the total error $\epsilon_{tot}.$ Hence, in order to get a meaningful binning of the degrees for each number of bins, we have built 
We propose a hierarchical clustering approach that takes into account the two   observations above  and tries to ``uniformly'' 
distribute  the total error in each bin. 
Hence, we cluster together consecutive degrees based on their maximum relative distance  and on the sum of their probabilities. 
The latter should be as close as possible to the threshold $1 /n,$ where $n$ is the number of  bins in the partition. 
This ensures that no bin   contributes disproportionately to the  error due to a large probability of the degrees in the bin.

In a bottom-up fashion,  we initially consider each degree to form an individual bin and iteratively join  those two  bins having the smallest distance (as defined below) until a single bin that contains all degrees is obtained. 
Hence, as a result we get a partition $B_n$ for all possible numbers of bins $n$. Note   that this computation is very cheap
compared to the solution of the ODEs for a fixed partition.
Typically, in hierarchical clustering, measures for the distance between two clusters (bins) are based on distance measures between
individual points (degrees) and on the choice of a linkage type. In the following, we propose a distance measure that can be directly applied to the bins and takes into account the degree distribution and the bin degrees' homogeneity.
To avoid bins of zero probability (all degrees in the bin have probability
zero), we assume that the distance of two bins $b_i$ and $b_j$, $i\neq j$ is zero if the probability of either of the two bins is zero.
Hence, those bins  will be merged with other bins. 
Otherwise, we define the distance $\text{dist}(b_i, b_j) $   between   two bins $b_i$ and $b_j$ as a linear combination of their probability and a homogeneity
measure of the union  $b_i \cup b_j$, i.e.,  
%\vw{I would like to rename $\text{dist}_{hmg}$. Does it appear later on again?} \todo{ok, no it does not}
\begin{align}\label{distMeasure}
\begin{split}
\text{dist}(b_i, b_j) &= P(b_i) + P(b_j) + \alpha \;\text{dist}_{hmg}(b_i, b_j),  \\ 
\text{dist}_{hmg}(b_i, b_j) & =  \frac{\big| \cavgk_{b_i}  - \cavgk_{b_j} \big|}{\big(\cavgk_{b_i} + \cavgk_{b_j} \big) / 2},
\end{split}
\end{align}
where $\text{dist}_{hmg}(b_i, b_j) $ is a homogeneity measure. Here, $\cavgk_{b_i}$ is the arithmetic mean
 of bin $b_i$, that is  $\cavgk_{b_i} = \frac{1}{|b_i|} \sum_{k \in b_i} k$. 
 Hence the formula measures the distance of the  means of two bins
 after normalizing it with the average of the two means. 
 This compensates for the effect of  large differences caused by  bins containing large degrees.
Finally,  $\alpha \in \mathbb{R}_{\geq 0}$ is used to scale   $\text{dist}_{hmg}$, since  typically $\text{dist}_{hmg}$ yields larger values than $P(b_i) + P(b_j)$. We empirically set $\alpha=0.2$; however,   
  the resulting binning turns out to be quite robust to small  variations of $\alpha$. 
%We stress that by defining directly the distance between two bins, we do not need to use any of the typical types of linkage for agglomerative clustering, thus avoiding the shortcomings of each of them.More importantly, we stress that using an hierarchical approach we do need not to predefine the number of bins, thus requiring to run the clustering only once to get the resulted binning for any number of bins. 

%\subsection{Choosing the number of bins}
After computing partitions $B_n$ for every possible bin number $n$, we choose an adequate $n$ giving a good tradeoff between accuracy and  computational time for the solution of the lumped equations.
Since we cannot measure the error $\epsilon(n):=\epsilon_{tot,B_n}$ between the lumped and the original equations directly, we resort to a simple heuristic to 
identify a suitable number of bins. 
This heuristic is based on the function
$$F(n)= \max_{t} ||\mathbf{x}(t, n) - \mathbf{x}(t, n-{j}^*)||,$$ where 
$\mathbf{x}(t,n)$, $\mathbf{x}(t,n-j^*)$ are the vectors containing the fraction of nodes in each state at time $t$, for the partitions  $B_n, B_{n-j^*}$, respectively. The step size ${j}^*\geq 1$ allows us to regulate how many steps (bins) 
are between two solutions we compare. In our implementation, we set ${j}^*=5$, to obtain a smoother error function and to  more easily detect the convergence of $F(n)$.
The basic idea of the heuristic is to increase $n$ and monitor the function $F(n)$, stopping when we observe no noticable variation of its value.  
%We remark that in our implementation, we compute $F(n)$ as the difference of two solutions 
%that are five steps apart to obtain a smoother error function and allow for easier detection of convergence.
Although the partitions that we compute are refinements of each other for increasing $n$,
we cannot guarantee that $\epsilon(n)$ and $F(n)$ are always monotonically decreasing.
However, $\epsilon(n)$ eventually approaches zero and 
for all examples that we considered, small changes of $F(n)$ for a number of steps indicated 
that with refinements of the corresponding partition $B_n$, only very small improvements were possible
in terms of $\epsilon(n)$.
%
%  between the solutions of the lumped equations of $n$ and $n-1$ bins will tend to zero. \vw{Yes, but not vice versa!
%  Shouldn't the argument be: If the improvement is only little, we stop increasing the number of bins.}
%The opposite is also true, as the sequence $\epsilon(n)$ is Cauchy  and will eventually stabilize to zero.
%\vw{The argument does not help in my view: there might be an earlier point where the elements of the sequence are
%close although we have not reached the point from which on \emph{all} following elements are close} 
%Just choose $n$ equal to $\maxk$.  \vw{??? I do not see why for this choice we have reached this point.}
In Fig.\,\ref{fig:sir} (b) and (d)   we see that   $F(n)$ is roughly proportional to  
$\epsilon(n)$ and hence the difference $ F(n) - F(n-j)$ 
can be used to identify the number $n$ beyond which  $\epsilon(n)$ decreases only very slowly. 
This proportionality between $F(n)$ and $\epsilon(n)$ was consistent for all models we have studied (see Fig.\,\ref{fig:sir} (b) and (d)). %\todo{shall we say about convexity as well? we need it actually}

The above observation gives rise to the heuristic described in Algorithm~\ref{alg:chooseBinsAlg}  to decide the number of bins for the lumped equations based on a gradient descent approach.
In the algorithm,
 $j\in\{1,\ldots,n-1\}$ is a fixed (limited) lookahead used to approximate the gradient of $F(n)$, and it is not necessarily equal to the step gap ${j}^*$ used in the computation of $F(n)$, even if in our implementation we used the same value, $j=5$, since smaller values (e.g.\ $j=1$) might not be as effective in indicating convergence.
Apart from the model, the network degree distribution and the   lookahead $j$, 
Algorithm~\ref{alg:chooseBinsAlg}  takes as input 
a predefined threshold $\delta$  
and a parameter $\gamma_n$ which is used for updating the number of bins for the next iteration. 
Note that increasing $n$ by a fixed number of bins in each step may lead to unnecessarily long running times.
When $n$ is small, an increase of the number of bins often gives a large improvement, 
 and hence $-\nabla F(n)$ is large as well. In that case, 
 a large increase of $n$ is  adequate.  
 In general, $\gamma_n$ can be either fixed for every iteration, i.e., $\gamma_n = \gamma$, or it can adaptively be chosen by a backtracking line search method \cite{wright1999}. In our implementation, we fixed $\delta = 10^{-3}$ and heuristically
 chose a fixed scaling constant $\gamma = 2 \cdot 10^3.$ 
 % as we need to search the space of integer numbers. 

\begin{table}
 \begin{algorithm}[H]
\begin{algorithmic}[1]
	\State $n \gets j+j^*+1$ 
	\State  Solve the lumped equations for all partitions necessary  to compute $F(n)$ and $F(n-j)$
	\State $\nabla F(n) = F(n) - F(n-j)$ 	
	 \If {$\nabla F(n) \leq \delta$}
			\State \Return $n.$
		\Else  \State $n \gets n + \max(1,\mathsf{floor}(- \gamma_n \nabla F(n)))$ and \textbf{goto} 2 
	\EndIf
\end{algorithmic}	
\caption{(Heuristic to choose the number of bins) \label{alg:chooseBinsAlg} } 
\end{algorithm}
\end{table}

\subsection{Results}
We test our lumping approach for two   well-studied models, namely the SIR model and a rumor spreading scenario, as well as 
for a third more complex system. 
For the simple SIR model, we tested  our method for different degree distributions and 
investigated how the running time of the method increases when  we vary the maximum degree $\maxk$ of the network. 
For the rumor spreading model, we show that even with a small number of bins the method is able to capture contact processes with more than one contact rules. 
Finally, using the complex disease model, we verify that the method performs  well even in cases of networks with many 
possible node states. 
For all models, we compare the dynamics and the running times of the original equations versus the lumped equations for a number of bins needed to get an error of at most $3\cdot 10^{-3}$. Note that for each model we chose, whenever possible, parameter values s.t.\ the original DBMF equations differ significantly from the PA equations.
In addition, we report the number of bins that was returned considering the stopping criterion of Algorithm~\ref{alg:chooseBinsAlg}.
In well, we provide a user-friendly python tool, called \emph{LUMPY} \cite{tool}, that takes a simple description of a contact process as input and   a degree distribution.
It automatically generates  all kinds of equations, both the full ones and the lumped ones, using the binning   heuristic discussed above. %Hence,  users can run their own contact processes and also compare the lumped with the original ODEs. 

%of the PA equations that significantly differ from those of the DBMF

\paragraph{\textbf{SIR Model}}

\begin{figure}[h!]
	\centering
	\subfigure[]{
		\includegraphics[width=0.45\linewidth]{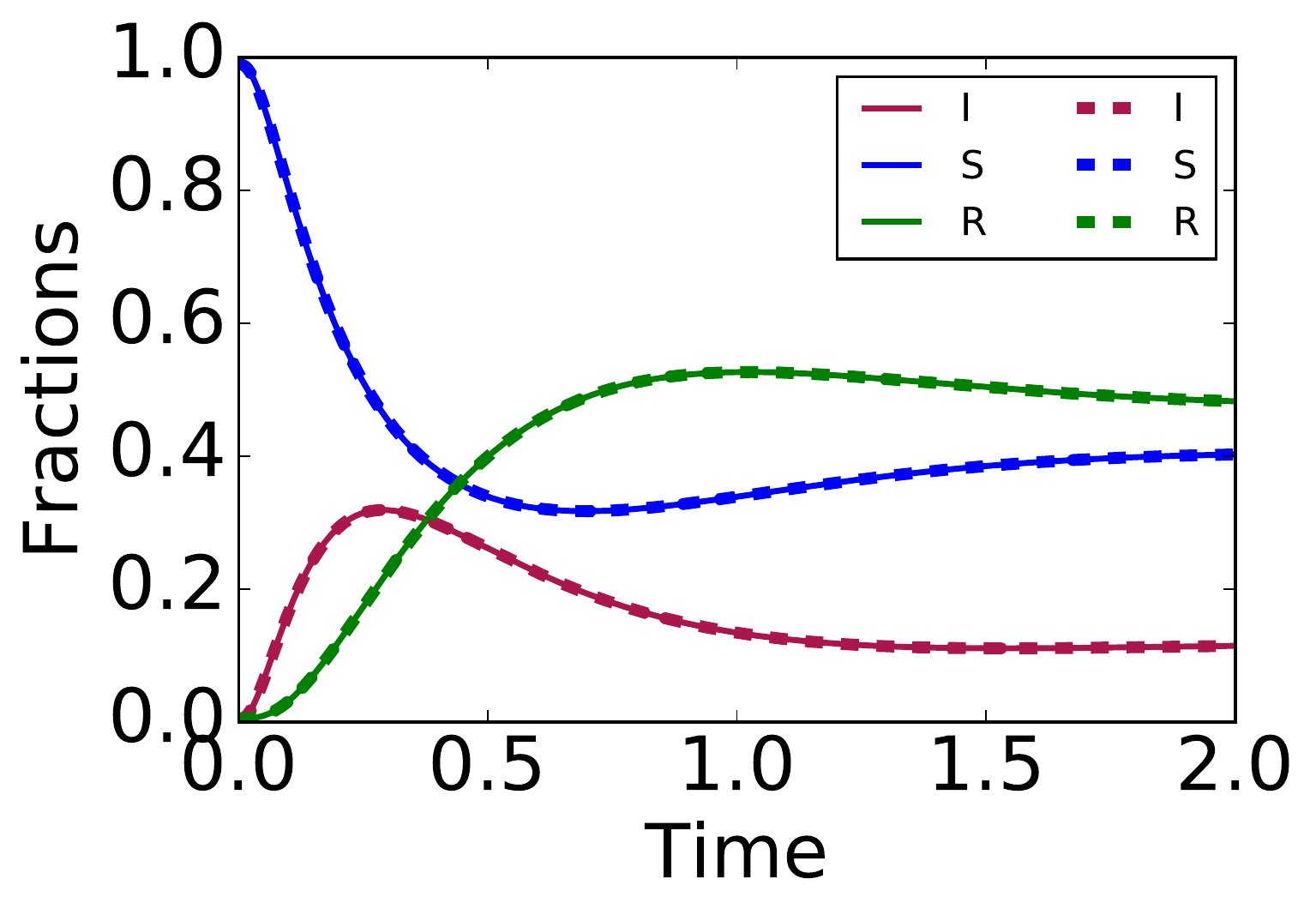} 
		\label{fig:sir_dbmf_traj}
	}
	\subfigure[]{
		\includegraphics[width=0.45\linewidth]{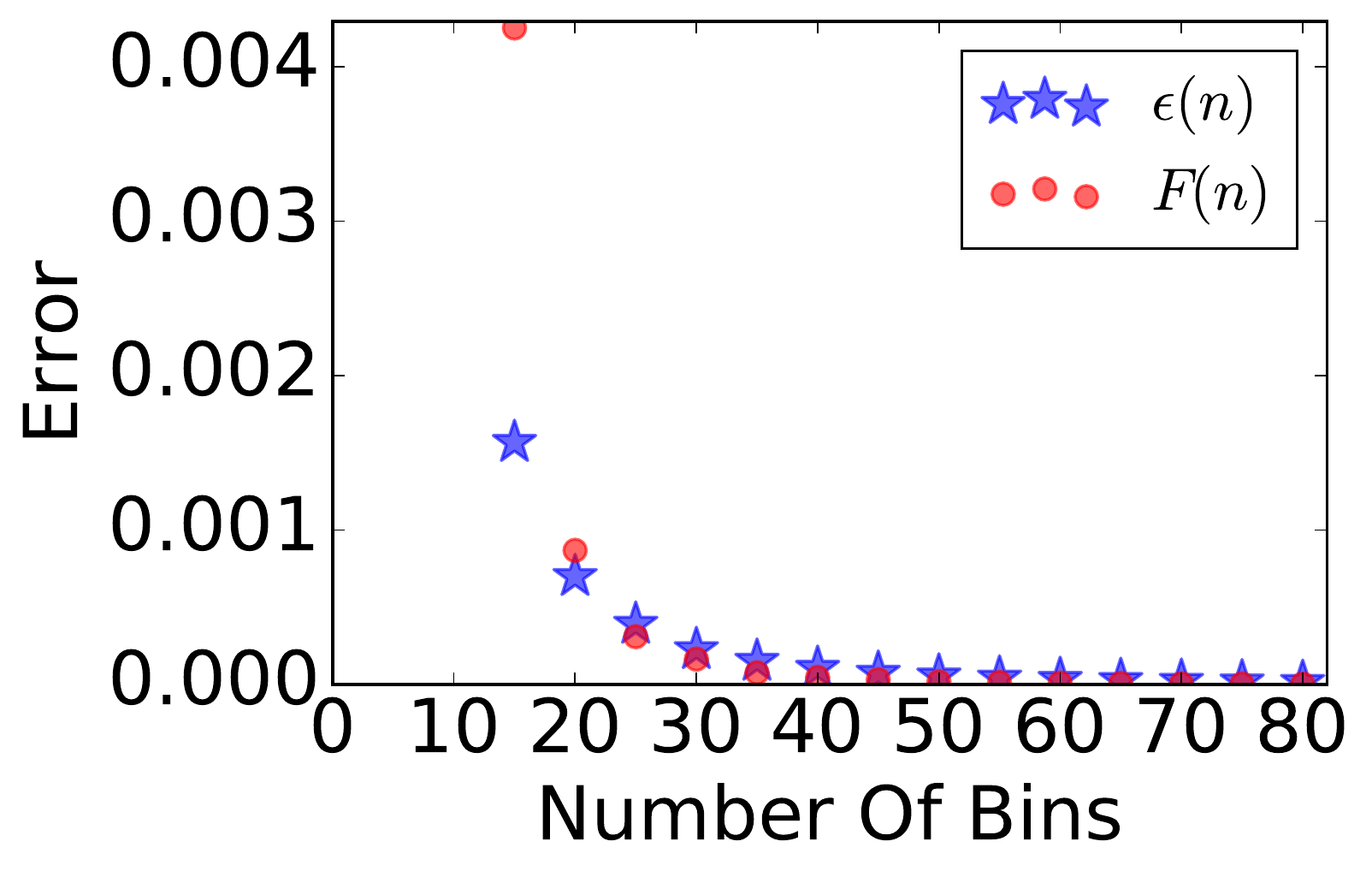} 
		\label{fig:sir_dbmf_errors}
	}
	\subfigure[]{ 
		\includegraphics[width=0.45\linewidth]{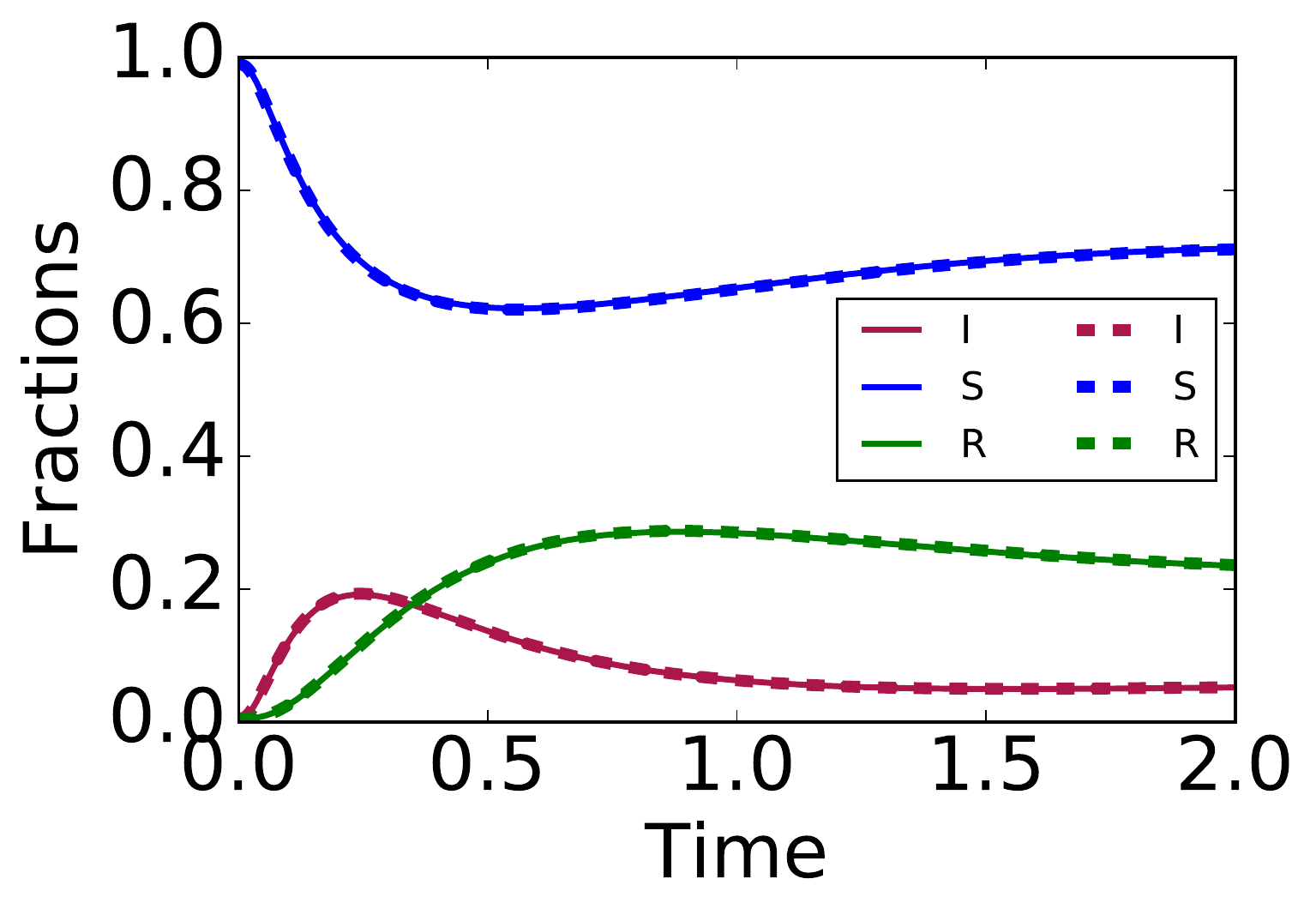} 
		\label{fig:sir_pa_traj}
	}
	\subfigure[]{ 
		\includegraphics[width=0.45\linewidth]{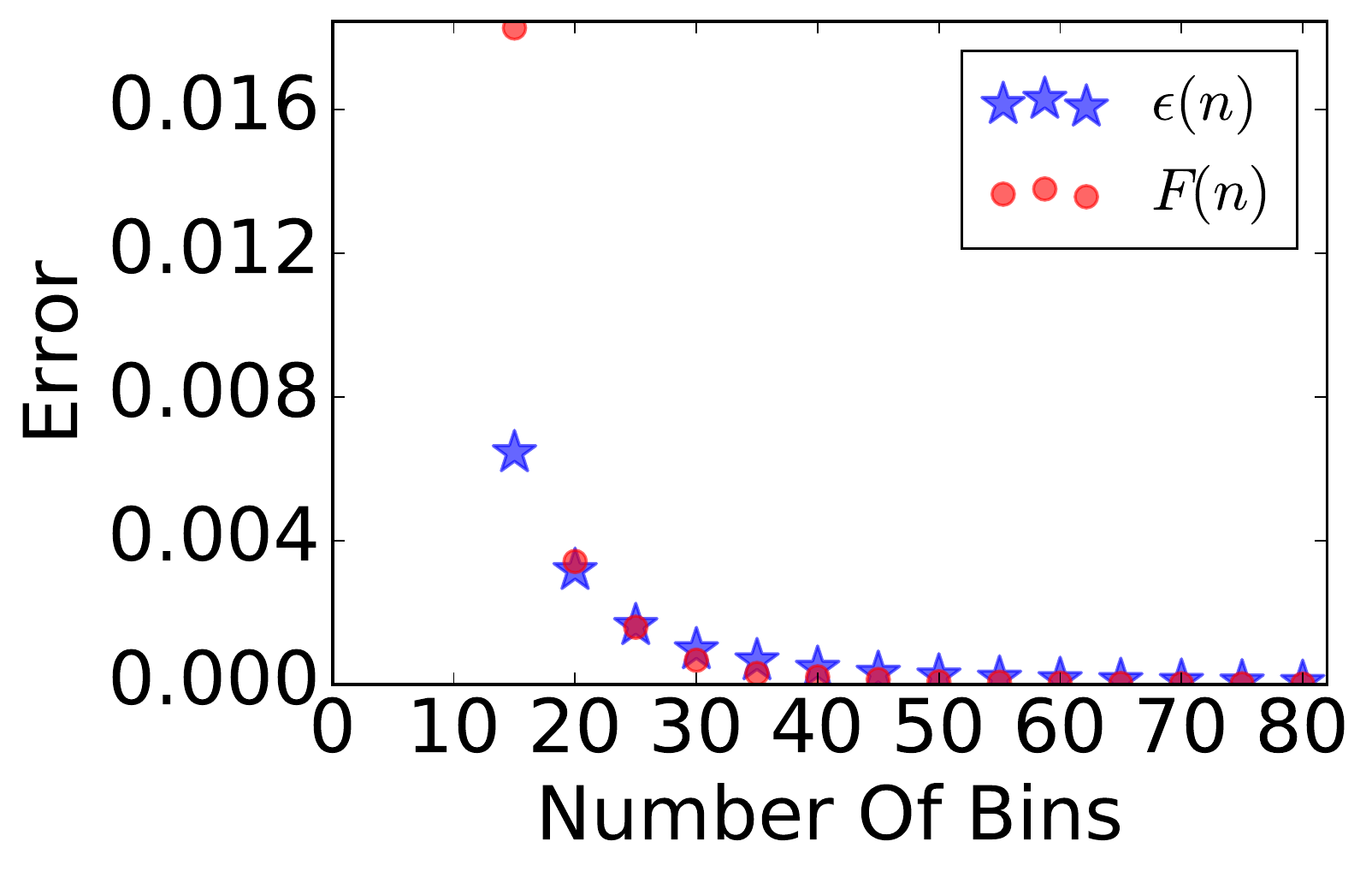} 
		\label{fig:sir_pa_errors}
	}
	\caption{SIR model: Total fraction of infected $(I)$, susceptible $(S)$ and recovered $(R)$ nodes based on the original (solid lines) vs.\ the lumped (dashed lines) DBMF (a) and PA (c) equations for a power law network with $\maxk = 10^3, P(k) \propto k^{-2.4}$. 
	The maximum error $\epsilon(n)$  between the full and the aggregated equations w.r.t.\ the number of  bins ($\star$) and the maximum difference $F(n)$ between the aggregated solutions for two consecutive bin numbers ($\circ$)
	are shown for the DBMF (b) and the PA (d) lumping, respectively. 
	\label{fig:sir}}
	\end{figure}

The first model we examined is the   SIR model, which consists of the following rules:
a susceptible node $S$ that interacts with an infected node gets infected at rate $\theta_1$; an infected node can recover at rate $\theta_2$;  and a recovered node can go back to the susceptible state at rate $\theta_3$:
\begin{alignat*}{3} 
& S + I  \;&\xrightarrow{\theta_1}& \;\;2I    \\
& I  \;&\xrightarrow{\theta_2}&\;\; R\\
& R \;&\xrightarrow{\theta_3}&\;\; S.
\end{alignat*}
Here for all the experiments we set $\theta = (\theta_1, \theta_2, \theta_3) = (6, 4, 1)$ and we assume the underlying network's structure follows a power law distribution with $\maxk = 10^3 $ and $P(k) \propto k^{-2.4}$.
In Fig.\,\ref{fig:sir_dbmf_traj} we show the comparison between the dynamics of the aggregated DBMF equations using 20 bins and the original DBMF equations and in Fig.\,\ref{fig:sir_pa_traj}  we compare the aggregated PA equations for 21 bins and the original PA equations. 
We see that it in both cases is almost impossible to distinguish between the global measures derived from the aggregated ODEs and those obtained from the original system of equations. 
In Fig.\,\ref{fig:sir_dbmf_errors}  and \ref{fig:sir_pa_errors} we plot the maximum error $\epsilon(n)$ between the full and the lumped DBMF and PA equations, respectively, as a function of the number of bins $n$ as well as $F(n)$, i.e., the maximum difference between the aggregated solutions for $n$ and $n-j$ ($j=5$).  Note that both errors quickly converge to zero as the number of bins increases.

\begin{figure}[h!]
	\centering

	\subfigure[]{ 
		\includegraphics[width=0.45\linewidth]{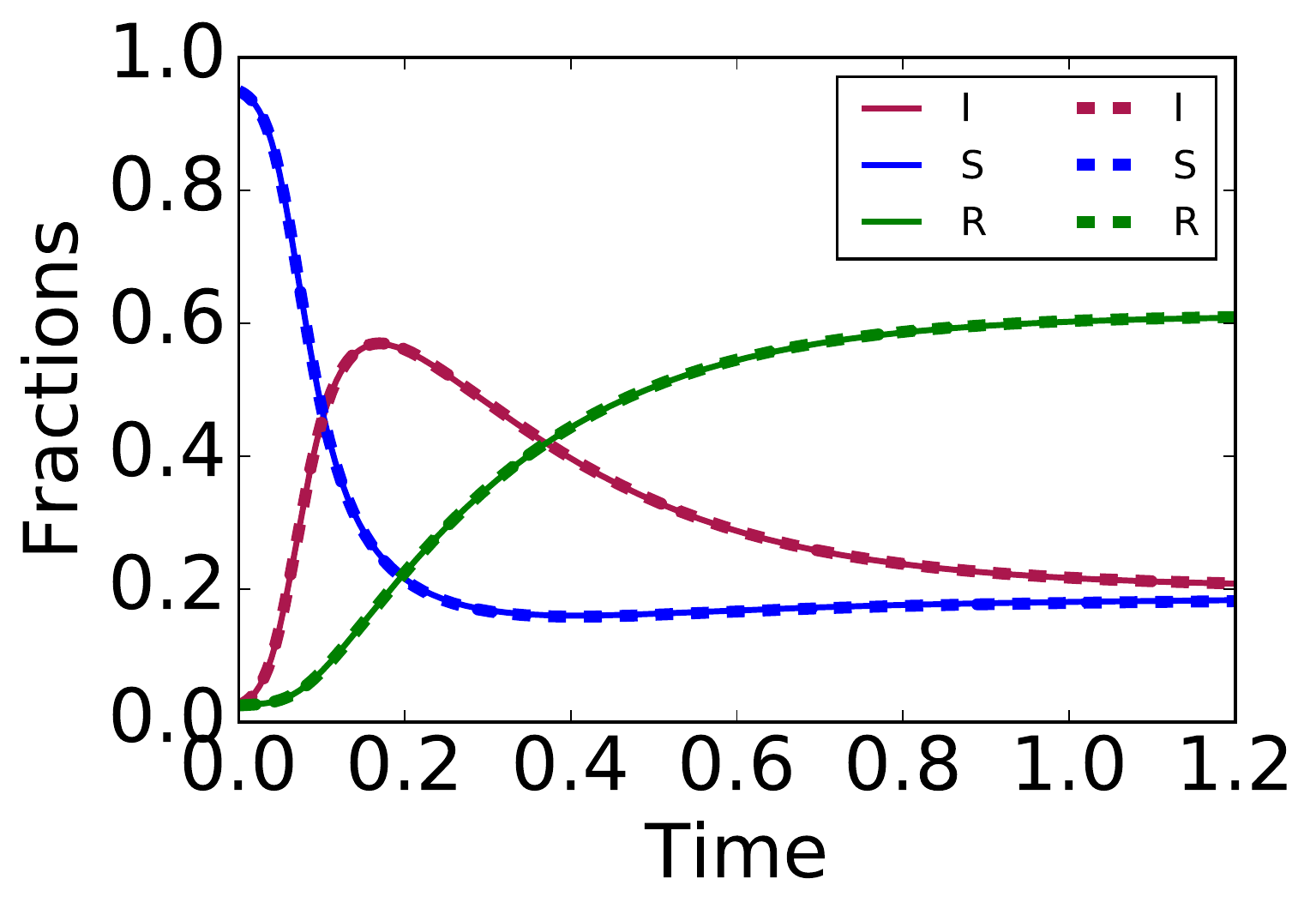} 
		\label{fig:sir_pa_uni_traj}
	}
	\subfigure[]{ 
		\includegraphics[width=0.45\linewidth]{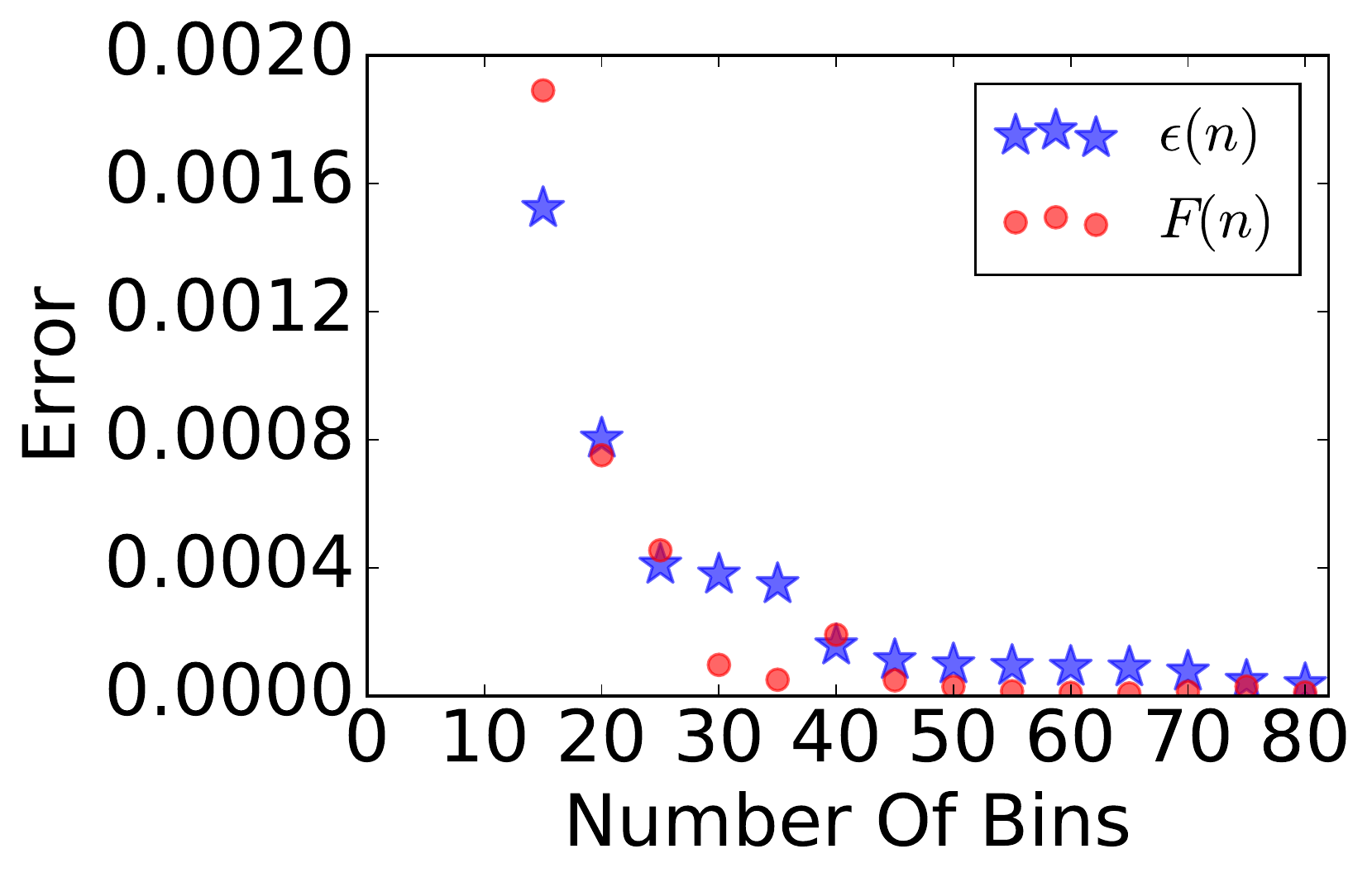} 
		\label{fig:sir_pa_uni_errors}
	}

	\caption{SIR model: (a): Total fraction of infected $(I)$, susceptible $(S)$ and recovered $(R)$ nodes based on the original (solid lines) vs.\ the lumped (dashed lines) PA equations for a uniform network with $\maxk = 500, P(k) = 1 /\maxk.$
		(b): The maximum error $\epsilon(n)$  between the full and the aggregated equations w.r.t.\!\!\, the number of  bins ($\star$) and the maximum difference $F(n)$ between the aggregated solutions for two consecutive bin numbers ($\circ$) for the PA lumping. 
		\label{fig:sir_uni}}
\end{figure}

In Fig.\,\ref{fig:sir_pa_uni_traj} and \ref{fig:sir_pa_uni_errors} we show  the same  plots  for a network with uniform degree distribution $P(k) = 1 /\maxk, $ where $\maxk = 500$, and  the solution of the PA equations (aggregated vs.\ original ones). For the chosen network size the dynamics of DBMF equations completely agree with those of PA.
Even though the uniform degree distribution is unrealistic for a typical contact network, we see that the lumped equations
   give very accurate results for a small number of bins. For an error $\epsilon(n)$ of at most $3\cdot 10^{-3}$ we only need $n=11$ bins and $132$ ODEs, in comparison to $500$ degrees and $6000$ ODEs of the full PA equations for this network. 
Note that the approximation works that accurately here, although due to the uniform assumption only the second part of the distance function in Eq.\,\eqref{distMeasure} (homogeneity of degrees)  is considered in clustering. 

\begin{figure}[h!]
	\centering
	\subfigure[]{ 
		\includegraphics[width=0.45\linewidth]{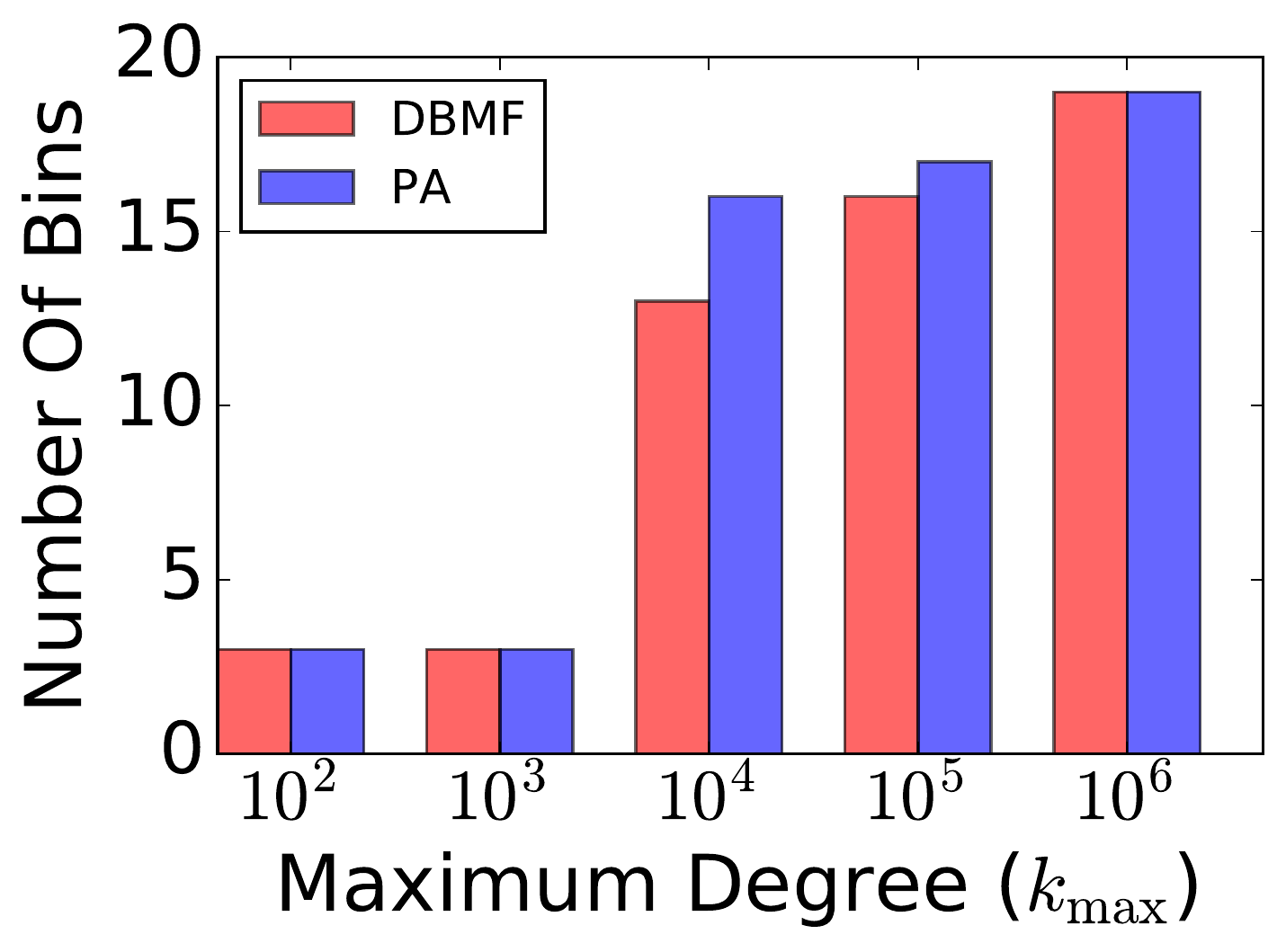}
		\label{fig:sir_bins} 
	}
	\subfigure[]{ 
		\includegraphics[width=0.45\linewidth]{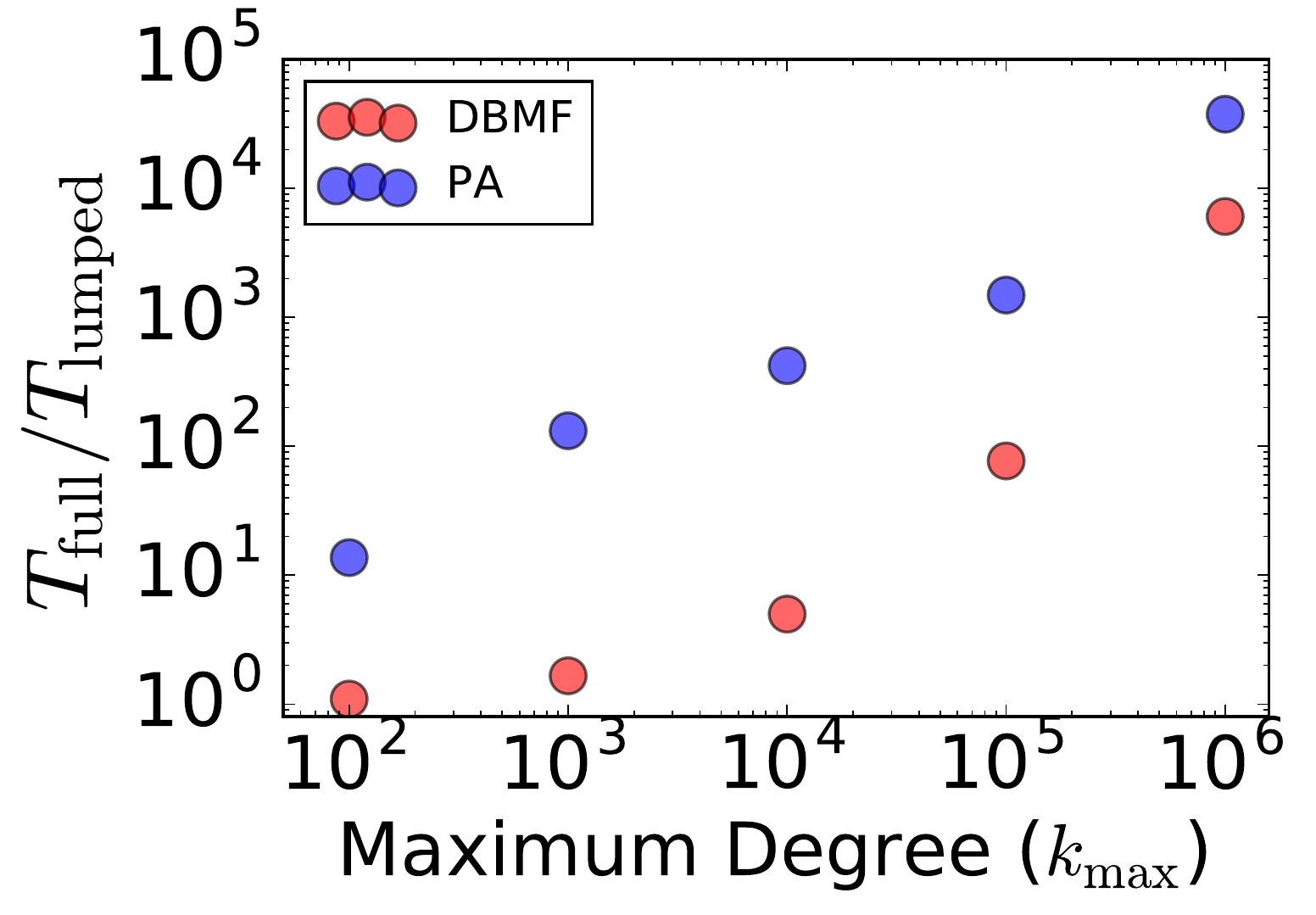}
		\label{fig:sir_times} 
	}
	\caption{SIR model: The number of bins needed for an approximation error of at most 
	$3 \cdot 10^{-3}$  for the solution of the DBMF and PA equations and for varying $\maxk$ (a), and the gain ratio of the computation times $T_{\mathrm{full}} / T_{\mathrm{lumped}}$ of lumped DBMF and PA equations for different $\maxk$ (b).	\label{fig:sir_bins_times}
} 
\end{figure}

In Fig.\,\ref{fig:sir_bins} we plot the number of bins needed for the solution of the DBMF and PA equations for 
an approximation error of at most $3\cdot 10^{-3}$  in a power law network with varying $\maxk.$ 
The network size grows exponentially, while the number of bins needed grows only linearly.
It is worth mentioning that in order to be sure that the dynamics of the network will change as $\maxk$ increases, we have changed the parameter $\alpha$ of the power law in such a way   that the coefficient of variation, i.e., $\sigma /\!\avgk$, of the distribution remains constant for different $\maxk$ (see also Appendix, \ref{dynamics}).

In Fig.\,\ref{fig:sir_times} we plot the gain ratio $T_{full} / T_{lumped}$ of the 
computation times for solving the aggregated equations instead of the full system. 
Generally the speedups are impressively high, and we see a significant  increase of the ratio as the network size increases. 
In particular, for $\maxk = 10^6$ we need $5.2$ days for solving the full equations, versus $12$ sec.\! for the lumped equations. 
The precise computation times of the needed lumped equations for varying $\maxk$ are listed in Table~\ref{tab:runningTimes}, where the additional time needed for the hierarchical clustering is in the order of seconds up to network size $\maxk = 10^4$ and in the order of minutes up to $\maxk = 10^6.$ 
%\todo{maybe here an asterisk about the greedy clustering}

In Table~\ref{tab:numOfBins} we include the number of bins returned by Algorithm~\ref{alg:chooseBinsAlg}, together with the corresponding error, for both DBMF and PA equations for varying $\maxk$. 
The number of bins returned by Algorithm~\ref{alg:chooseBinsAlg} for $\delta=10^{-3}$ yields very accurate approximations (error $\ll$ $3\cdot 10^{-3}$), as the chosen stopping criterion of the heuristic is rather conservative.
In any case, the speedups compared to a full  solution of the equations are very similar to those in Fig.\,\ref{fig:sir_times}.

\begin{table}[b]
	\caption{\label{tab:runningTimes}%
		SIR model: Running times (in seconds) of the full equations ($T_{\mathrm{full}}$) vs. running times of the lumped equations ($T_{\mathrm{lumped}}$) for all network sizes.}
	\small
	\begin{ruledtabular}
		\begin{tabular}{lrrrrr}
			\textrm{$\maxk$}&
			\textrm{DBMF($T_{\mathrm{full}}$)}&
			\textrm{DBMF($T_{\mathrm{lumped}}$)}&
			\textrm{PA($T_{\mathrm{full}}$)}&
			\textrm{PA($T_{\mathrm{lumped}}$)}\\
			\colrule
			$10^2$ & 3 & 2  & 1   & 0.07  \\
			$10^3$ & 5 & 3 &  9  & 0.06 \\
			$10^4$ & 10  & 2 & 126  & 0.32  \\
			$10^5$ & 228 & 3 & 4428  & 3~~~~ \\
			$10^6$ & 18177  & 3  & 449280  & 12~~~ 
		\end{tabular}
	\end{ruledtabular}
\end{table}
%5.2 days
\begin{table}[b]
	\caption{\label{tab:numOfBins}%
		SIR model: Number of bins returned by Alg.~\ref{alg:chooseBinsAlg}. 
	}
\begin{ruledtabular}
	\begin{tabular}{lrrrrr}
		\textrm{$\maxk$}&
		\textrm{DBMF}&
		\textrm{DBMF $\epsilon$}&
		\textrm{PA}&
		\textrm{PA $\epsilon$}& \\
		\colrule
	    $10^2$ & 20 & $3.80\cdot 10^{-7}$& 20& $2.70 \cdot 10^{-8}$\\
		$10^3$ & 20 & $6.05 \cdot 10^{-5}$& 20& $7.95 \cdot 10^{-6}$\\
		$10^4$ & 55 & $1.04 \cdot 10^{-4}$& 56& $1.80 \cdot 10^{-4}$\\
		$10^5$ & 37 & $4.47 \cdot 10^{-4}$& 37& $4.52 \cdot 10^{-4}$\\
		$10^6$ & 35 & $5.87 \cdot 10^{-4}$ & 35& $6.18 \cdot 10^{-4}$
		\normalsize
	\end{tabular}
\end{ruledtabular}
\end{table}

\begin{figure}[h!]
	\centering
	\subfigure[]{ 
		\includegraphics[width=0.45\linewidth]{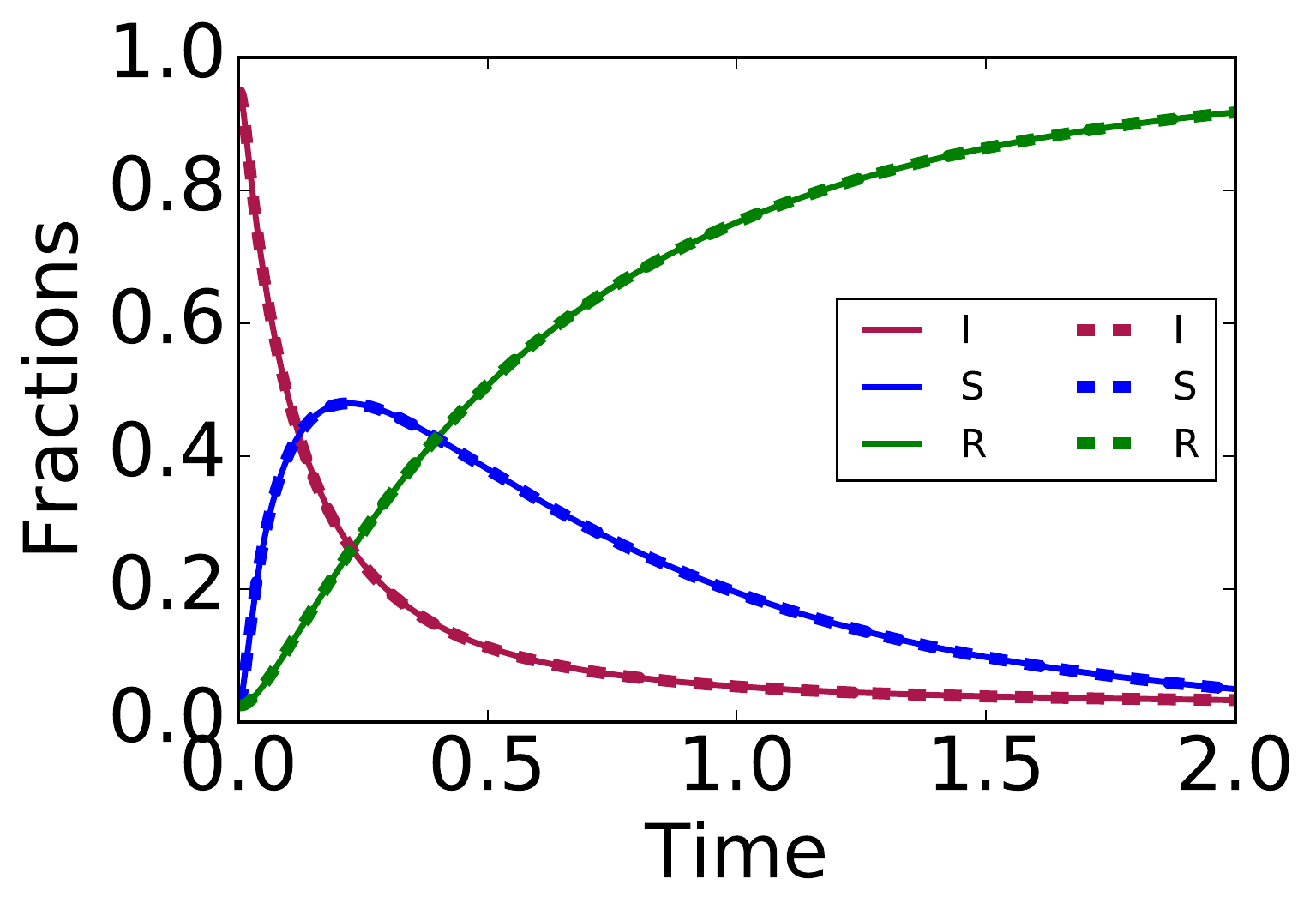}
		\label{fig:rumor_dbmf} 
	}
	\subfigure[]{ 
		\includegraphics[width=0.45\linewidth]{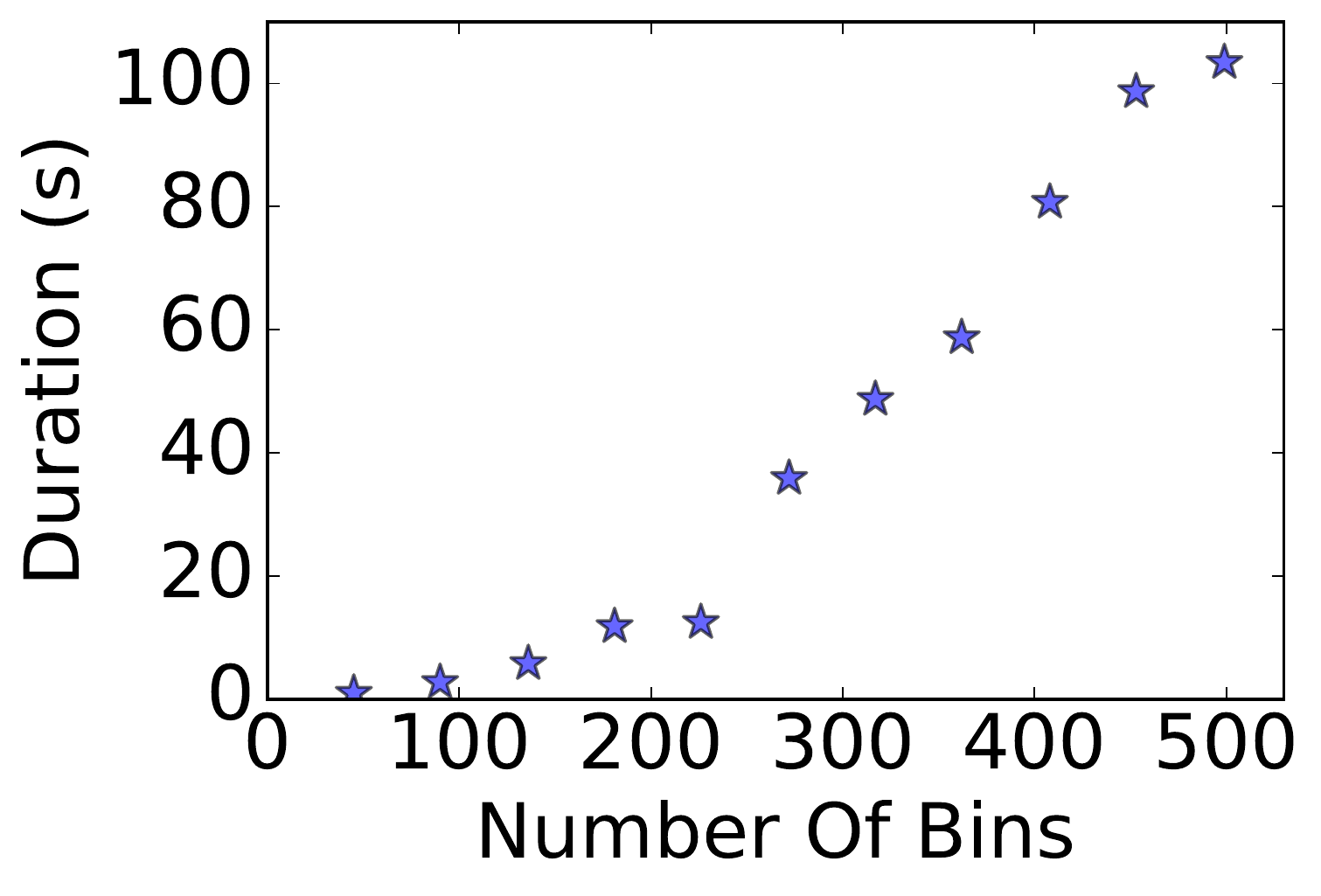}
		\label{fig:rumor_dbmf_times} 
	}
	\subfigure[]{ 
		\includegraphics[width=0.45\linewidth]{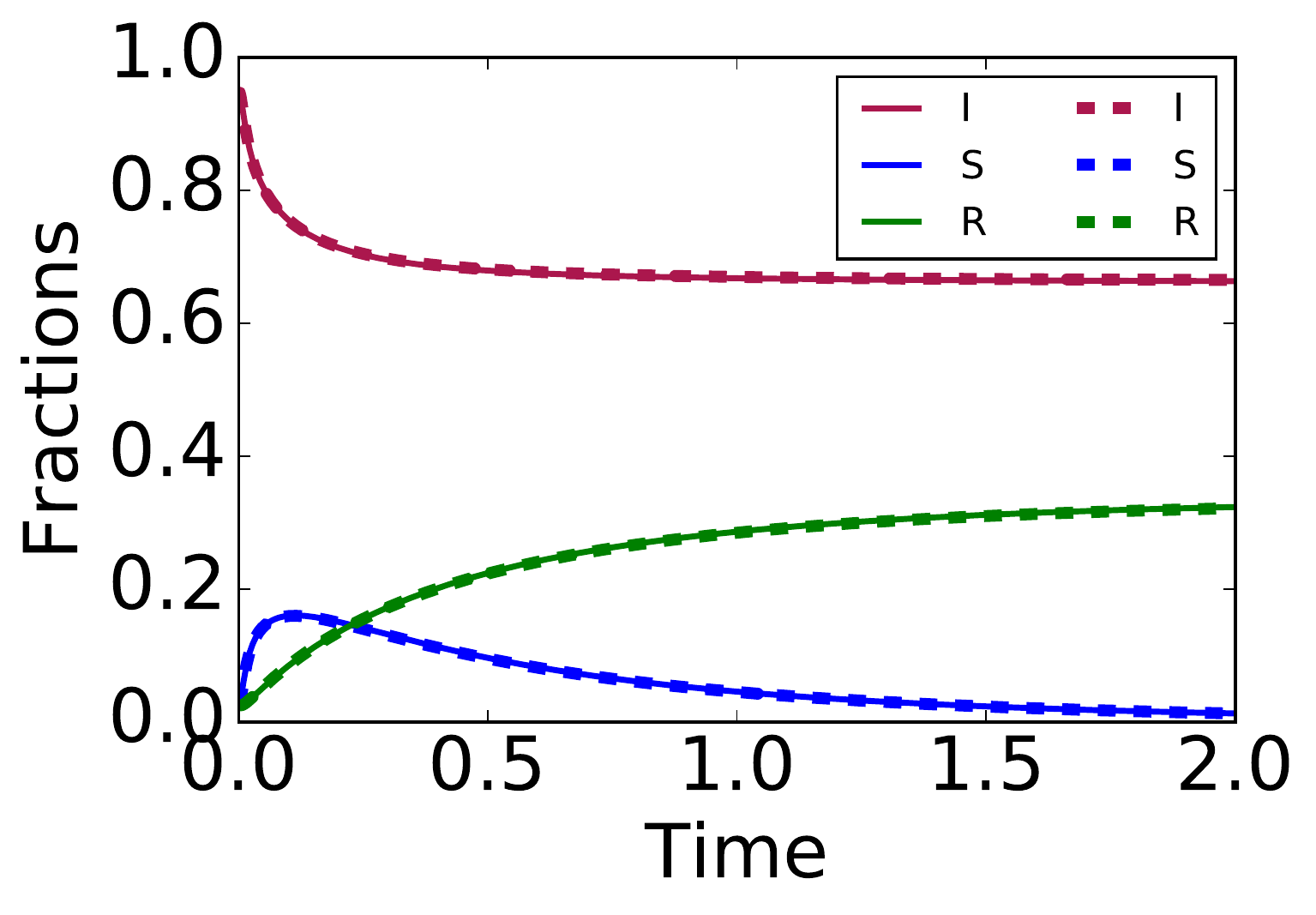}
		\label{fig:rumor_pa} 
	}
	\subfigure[]{ 
		\includegraphics[width=0.46\linewidth]{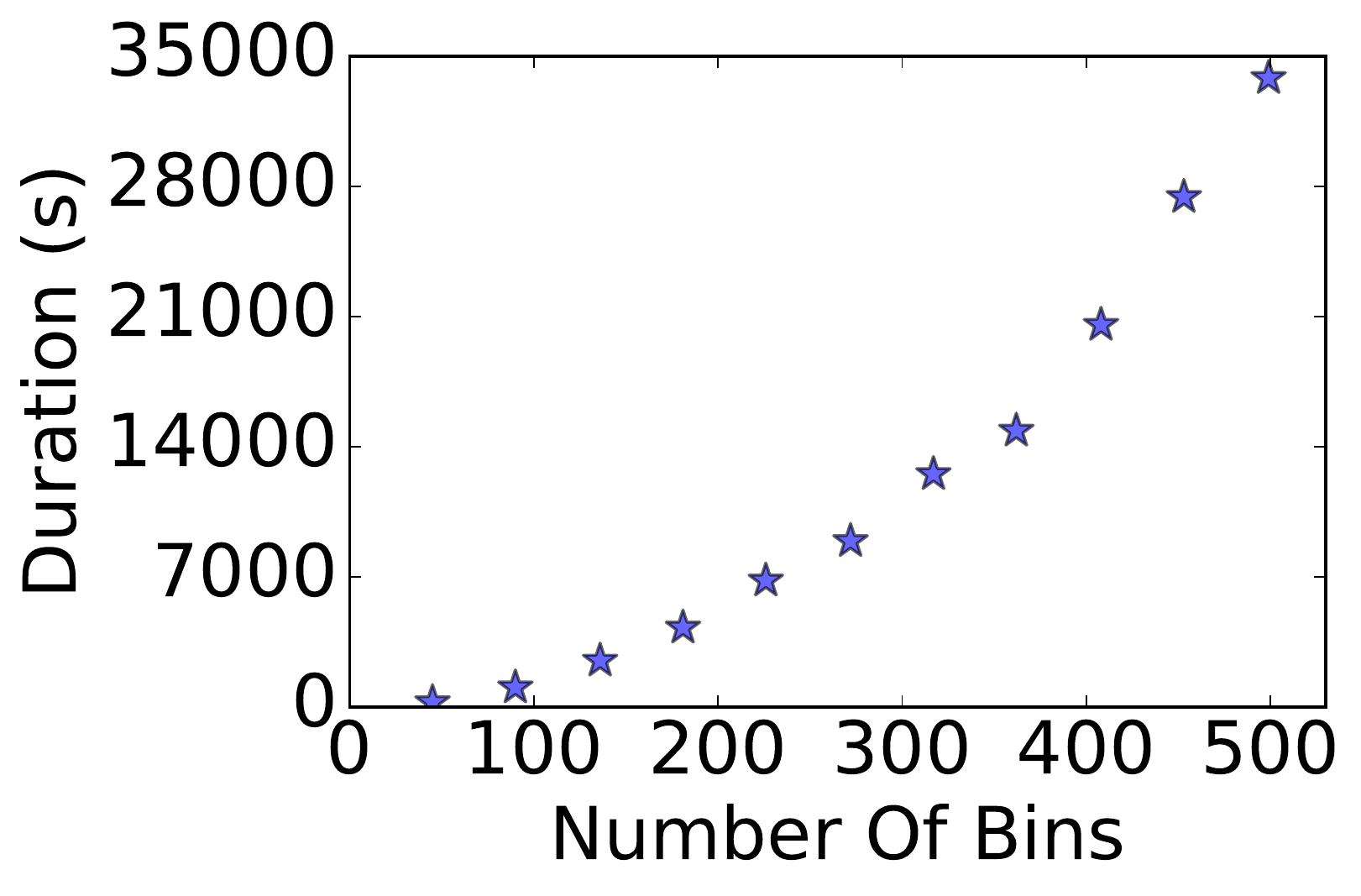}
		\label{fig:rumor_pa_times} 
	}
	\caption{Rumor spreading model: Total fraction of ignorant $(I)$, spreaders $(S)$ and stiflers $(R)$  in a power law network with $\maxk = 500$ for the original DBMF (a) and PA (c) equations (solid lines) and solution of the corresponding  lumped equations (dashed
	lines). 
	 $\star:$ The running time of the aggregated equations w.r.t.\,the number of bins used for DBMF (b) and PA (d). \label{fig:rumor_dbmf_pa_times}} 
\end{figure}

\paragraph{\textbf{Rumor Spreading Model}}
As a second model, we consider a rumor spreading model \cite{barrat2008} that includes three different types of individuals: ignorants (I), spreaders (S), and stiflers (R). 
An ignorant that is in contact with a spreader becomes a spreader  at rate $\lambda.$ 
A spreader becomes a stifler   when the spreader is in contact either with a spreader or a stifler; in both cases the
corresponding transition rate is $\alpha$: 
\begin{alignat*}{3} 
& S + I  \;&\xrightarrow{\lambda}& \;\;2S    \\
& S+R  \;&\xrightarrow{\alpha}&\;\; 2R\\
& 2S \;&\xrightarrow{\alpha}&\;\; S+R.
\end{alignat*}
Here, we set $\lambda = 6.0$ and $\alpha = 0.7$ and,
for the network's degree distribution, we consider a truncated power law $P(k) \propto k^{-2.2}$ with $\maxk = 500.$ 
In Fig.\,\ref{fig:rumor_dbmf} we show the dynamics of the full DBMF ODEs compared to the DBMF lumped equations
 and in Fig.\,\ref{fig:rumor_pa} the corresponding comparison for  the PA  equations. 
 We note that DBMF here gives  a poor approximation, which is qualitatively different from the PA solution.
Even though there is a great difference between the solutions of the two systems of equations, 
%and the particularly complex dynamics of the PA equations include two local maxima over time, 
the lumped equations approximate very accurately the original equations for both cases.
 For this model, we needed only 13 bins and 39 ODEs for the DBMF case and 27 bins and 324 ODEs for the PA  in order to get a maximum error of $3\cdot 10^{-3}$.
 The heuristic in Alg.\,\ref{alg:chooseBinsAlg} with $\delta=10^{-3}$ returned 43 bins with an error of $10^{-4}$ for the first and 60 bins with an error of $2 \cdot 10^{-4}$ for the second case.
% \gerrit{The heuristic yields 43 bins for DBMF and an error off 0.00010664215
% 	and 60 bins for PA with an error of 0.000218358787408.
% 	Non-heuristic is 13 and 27 bins.}
 The computation times for solving the lumped equations of DBMF and PA for every possible number of bins are shown in Fig.\,\ref{fig:rumor_dbmf_pa_times} (b) and (d). Again, we see huge computational gains when solving the lumped equations instead, especially in the   case of PA. 

\paragraph{\textbf{SIIIR Model}}
Recent infection models  describe the infection process in more detail by considering multiple infection phases \cite{miller2013}.  
In HIV, for instance, the early phase is well known to be significantly more infectious than a longer-lasting chronic phase \cite{pilcher2004, eames2002}. In order to capture such behavior, here we incorporate the  phases as different states of the individuals, i.e., $I, II, III,$ with different transmission rates:
\begin{alignat*}{6} 
& S + I  \;&\xrightarrow{\theta_1}& \;\;2I   \hspace{8pt}
& II\;&\xrightarrow{\theta_5}&\;\; III \hspace{8pt} \\
& S+ II  \;&\xrightarrow{\theta_2}&\;\; I + II
&III\;&\xrightarrow{\theta_6}&\;\; R \\
& S + III\;&\xrightarrow{\theta_3}&\;\; I+ III \hspace{8pt}
&R\;&\xrightarrow{\theta_7}&\;\; S \\
& I\;&\xrightarrow{\theta_4}&\;\; II 
\end{alignat*}

\begin{figure}[h!]
	\centering
	\subfigure[]{ 
		\includegraphics[width=0.45\linewidth]{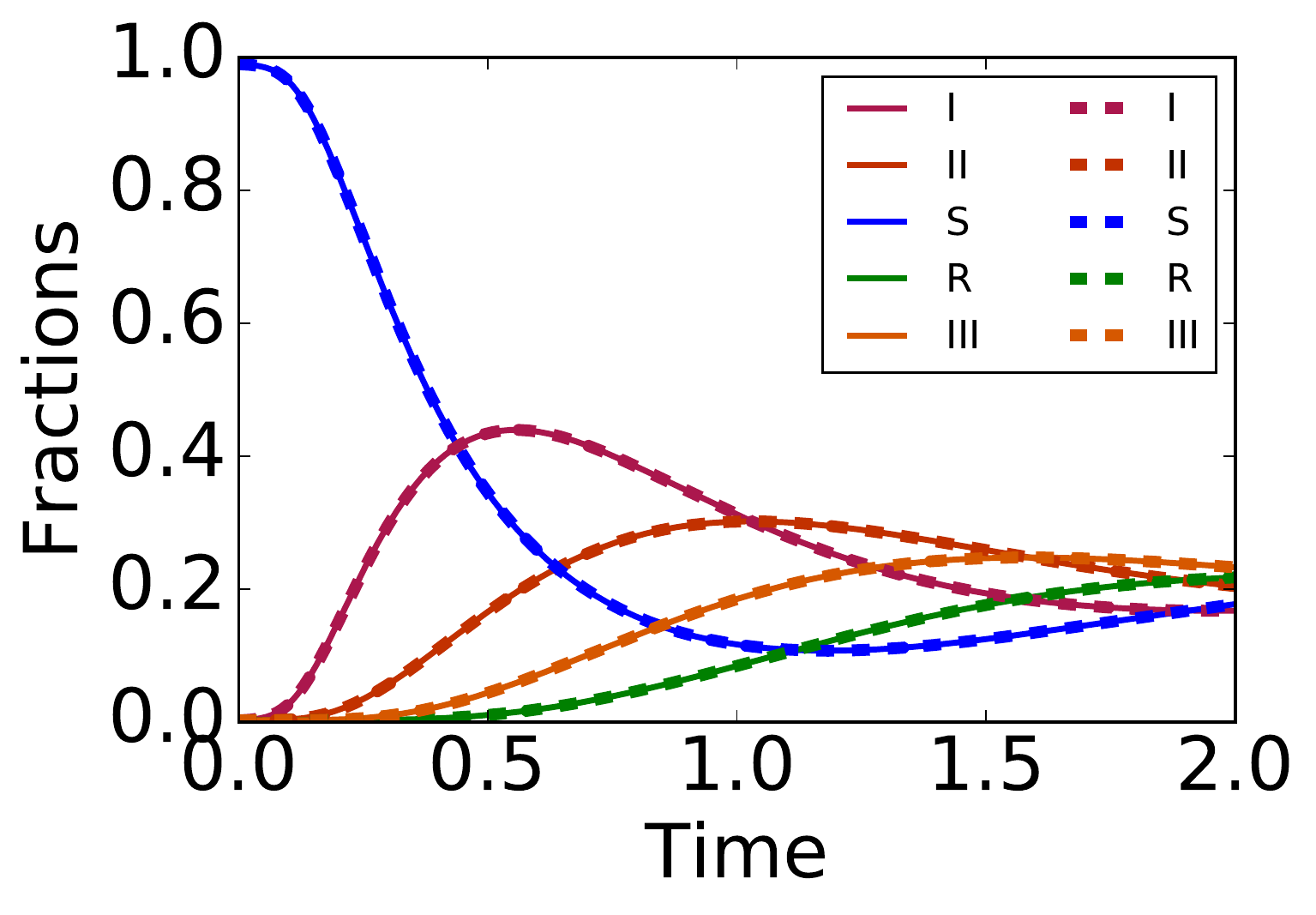}
		\label{fig:siiir_dbmf} 
	}
	\subfigure[]{ 
		\includegraphics[width=0.45\linewidth]{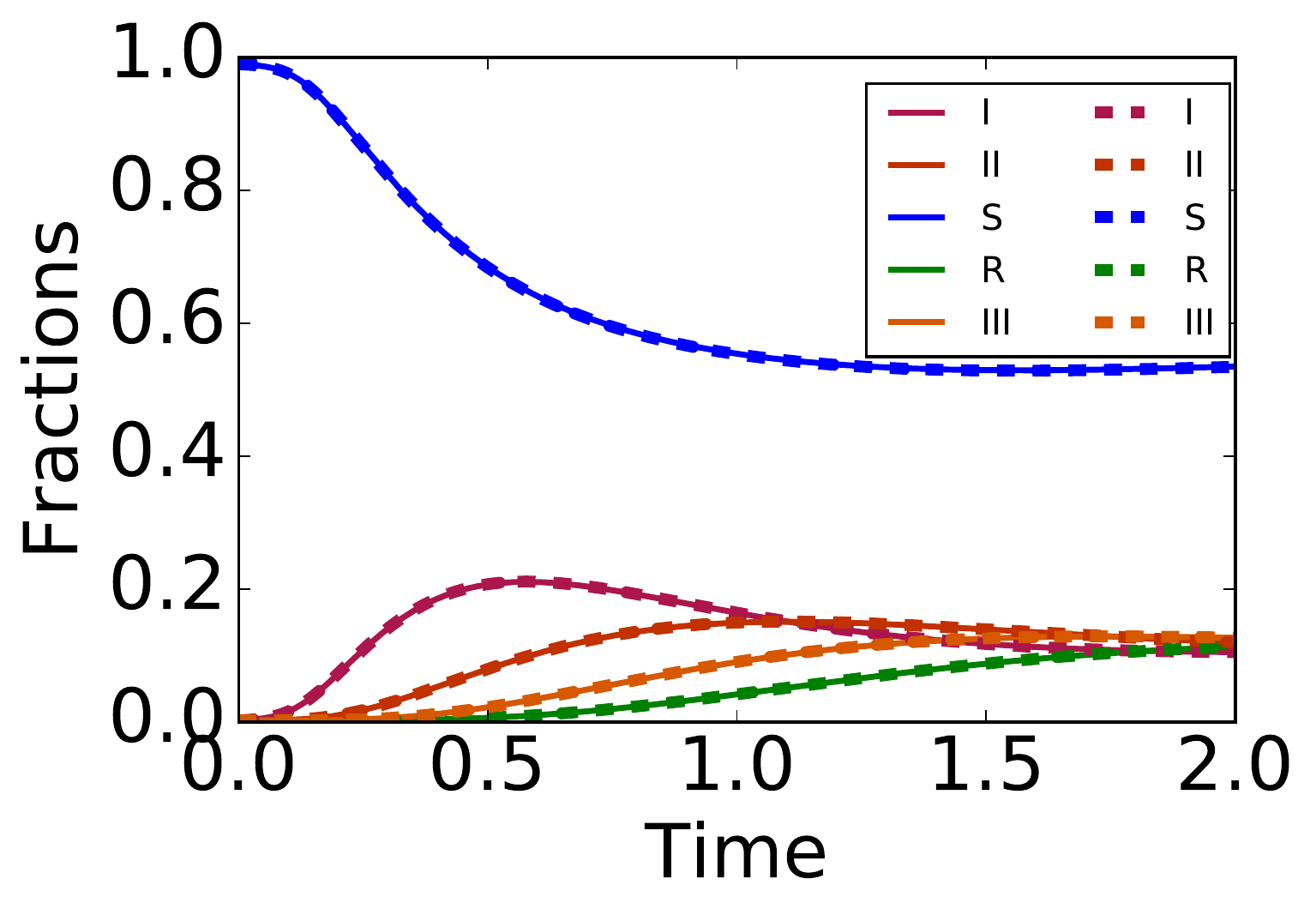}
		\label{fig:siiir_pa} 
	}
	\caption{SIIIR model: Total fraction of infected $(I), (II), (III)$ nodes of phase $I, II, III$, respectively, susceptible $(S)$ and recovered $(R)$ nodes in a power law network with $\maxk = 100, P(k) \propto k^{-2.5}$. 
	The solid lines show the solution of the original DBMF (a) and PA (b) equations, and the dashed lines the solution of the
	corresponding lumped DBMF (a) and PA (b) equations.
	% To get an approximation within a maximum error of $\leq $3\cdot 10^{-3}$$ using the lumping method we needed $15$ bins for both DBMF and PA equations.
	\label{fig:siiir_dbmf_pa}}
\end{figure}

We fix $\theta = (\theta_1, \ldots, \theta_7) = (5, 1.5, 1, 2, 2, 2, 2)$ and we test our lumped equations with the original DBMF and PA solutions taken for the scale-free network $(P(k) \propto k^{-2.5}, \maxk = 100)$  of sexual partners over a time period of one year \cite{liljeros2001}.
Despite the large number of possible node states, the dynamics of the total fractions of all states is captured very accurately (error of less than $3\cdot 10^{-3}$) by the lumped equations with 20 bins in both DBMF (Fig.\,\ref{fig:siiir_dbmf}) and PA (Fig.\,\ref{fig:siiir_pa}). 
For this system, Algorithm~\ref{alg:chooseBinsAlg} returned 39 bins for DBMF and 46 bins for PA with an error of $4.7 \cdot 10^{-5}$ and $5 \cdot 10^{-5}$, respectively, when $\delta=10^{-3}$ was chosen.
%\todo{gerrit?} \gerrit{PA is 17 and 60 bins and 1.84007997035e-05 error, DBMF is 15 and 54 bins and 1.80030984601e-05 error}
Last, it is worth mentioning that we also tested the robustness of our method regarding the choice of the initial fractions. 
In \cite{gleeson2011} and \cite{gleeson2013} the author uses for each state $s$ a constant initial fraction for all $x_{s,k}$ for every $k$, which is equal to the initial total fraction of $x_s.$ Here, we considered in addition completely random initial fractions for each $x_{s,k}$ for all states $s$, and this initialization returned an even smaller error function in the beginning (see also Appendix, \ref{initialCond}), mainly because we avoid the extreme values that might be present in our initialization, such as the total fraction of susceptible nodes, which is here close to one. Hence, we conclude that the method is robust against different initial conditions.

\FloatBarrier

\section{Conclusion} \label{sec:conclusions}

Studying complex dynamic processes happening on  networks is a daunting computational task when relying on simulation. 
To ameliorate this problem, several forms of approximations have been considered, all trying to take  the topology of the network into account to some extent: 
degree-based mean-field \cite{satoras2001}, AME, and pair approximation \cite{gleeson2013}. AME and PA equations, in particular, seem to capture with precision the dynamics of a  large variety of network topologies, but in  \cite{gleeson2013} they are typically restricted to the binary case, i.e., to processes with only two local states. 
Here, we consider the AME and the  subsequent derivation of the PA  in the general case of contact processes of $n$ states.  
The total number of equations for both DBMF and PA is directly dependent on the maximum degree, $|S| \maxk$ and $|S|^2 \maxk$, respectively, and hence the total computation time even for these approximate methods can be  demanding for large networks with many states. 
%
%
%DBMF equations can be quite accurate for dynamics on well connected networks \cite{dorogovtsev2008} but they are known to perform poorly on sparse networks. Trying to capture dynamics in such networks,
%Gleeson in \cite{gleeson2013} proposed for a binary state network AME, a system of odes which describe in addition the state of the neighbors of a $k$ degree node of a certain state.
%From this he was able to derive PA equations by imposing the binomial assumption to the number of neighbors that are in one of the two network's states. 
%AME and PA equations seem to capture the dynamics of a much larger variety of network topologies.
%Here we generalize the AME and we show subsequently how PA can be derived from it in the general case of contact processes of $n$ states.  The total number of equations for both DBMF and PA is directly dependent on the network's largest degree, $|S| \maxk$ and $|S|^2 \maxk,$ respectively and hence the total computational time can be particularly demanding in large networks with many states. 

In this paper, we argue that although the network structure is proven to be essential for its dynamics, we do not need a full resolution to get precise results for the time evolution of the total fractions.  For instance, it is expected that the degrees belonging to the tail of a power law should not significantly change the precision of the equations if   clustered together. 
Motivated by this insight, we investigated whether it is possible to capture the dynamics of DBMF and PA equations by solving  a smaller set of equations. 
Taking advantage of the  degree distribution characteristics, and additionally exploiting a notion of homogeneity of the differences between separate degrees, we provide a mechanism according to which we can effectively reduce the resolution of network degrees, clustering them into a small number of bins and constructing a small set of aggregated differential equations.

The results in terms of computational gain are particularly impressive:  we manage to approximate collective network statistics, namely the total fractions of nodes in each state, produced by DBMF and PA equations by solving a significantly smaller number of differential equations and gaining a computational speedup of several orders of magnitude. We are able to reproduce these results for a number of different network degree distributions and a number of different models.  

Furthermore, the procedure to construct the reduced set of equations is fully automatic, and it is implemented as 
an open-source Python tool \cite{tool}. 
The tool takes a compact description of the model as input and returns Python files containing the lumped and the full DBMF and PA equations as output. This makes   validation of the results and   experimentation with other models a very easy task.

The natural direction in which to extend this work is to provide a similar approach to aggregate the AME equations. A direct strategy would be to use the same binning as for PA equations and lump AME only over different degrees $k$. For each of the formed bins, we would then need to solve $|S| |\Vecm_{\maxkB}|$ equations. This is guaranteed to give at least as good results as the  lumped PA equations.  The quality, however, may depend on whether the solution of the original PA equations is close to the solution of the AME. However, getting a simultaneous aggregation over degrees $k \in K$ and neighbors $\vecm \in \Vecm_k$ which is close to the solution of full AME equations could give a tremendous computational boost and make AME tractable regardless of the network size.
Finally, a modification of the presented lumping method could successfully be applied to  adaptive networks, where the degree distribution changes smoothly over time. 
In this case, we expect that ``on-the-fly" and local rearrangements of the clusters could be sufficient to   capture  well the 
topology changes of such networks.

\begin{acknowledgments}
	We would like to thank Mayank Goyal for an early implementation of the binning approach on the binary state case dynamics as part of his master's thesis.
\end{acknowledgments}

\appendix

\section{State distribution of the neighborhood in uncorrelated networks}\label{app:A1}
In this section, we prove Eq.\,\eqref{eq:probs} for uncorrelated networks, i.e., 
$p_k[s'] = p[s'] = 1/{\avgk}\sum_{k' \in K}  {P(k') k' x_{s', k'}} $.
  
We consider first the probability $q_{k}(k')$ that, given a node of degree $k$, a randomly chosen direct neighbor of it has degree $k'$.
 Since in an uncorrelated network the above probability is independent of $k$, it is the same as the probability $q(k')$
 of choosing  a random edge that connects to a node of degree $k'$.
% connects a  node of degree $k'$ to some node with arbitrary degree. 
 Hence:
\begin{eqnarray}
\label{uncorellated}
&&  q_{k}(k') = q(k')  \nonumber \\[1ex]
&=& \frac{\#(\mathrm{edges \;connected\; to\;k'\;node})}{\#(\mathrm{edges\;connected \;to\;arbitrary \;node})} \nonumber \\[1ex]
&=& \frac{k' N_{k'}}{2 |E|} =  \frac{k' N_{k'}/ N}{2 |E|/N} = \frac{k' P(k') }{\avgk}, \nonumber \\
\end{eqnarray} \normalsize
where $N_{k'}$ is the number of nodes of degree $k'$ and $|E|$ is the total number of edges.
Please note here the subtle but important difference between $q (k')$, which is the probability of picking a $k'$ node after randomly choosing a network edge, and $P(k')$, which is the probability of a random node being of degree $k'$.
Thus the probability that a random neighbor of a node of degree $k$ in an uncorrelated network is in state $s'$ is independent of $k$,  and   given by:
\begin{eqnarray}
\label{numOfAME}
&& p_k[s'] = p[s'] = \sum_{k' \in K} x_{s',k'} q(k') \nonumber \\ 
&=& \sum_{k' \in K} x_{s',k'} \frac{k' P(k') }{\avgk}.  \nonumber 
\end{eqnarray} \normalsize

\section{Number of AME equations} \label{numberOfAME}
Without considering the redundant equations, the number of equations of the general AME case with $|S|$ states
is  $\binom{\maxk + |S|}{|S|-1} (\maxk + 1).$ \\

\small
\textbf{Proof}: 
\begin{eqnarray}
\label{numOfAME}
&& |S| \sum_{k=0}^{\maxk}  \binom{k+ |S| -1}{|S|-1}  = |S| \binom{\maxk + |S|}{|S|} \nonumber \\
&=&  |S| \frac{(\maxk + |S|)!}{|S|! \;\maxk!} = \frac{(\maxk + |S|)!}{(|S| - 1)! \; (\maxk + 1)!} (\maxk + 1) \nonumber \\
&=& \binom{\maxk + |S|}{|S|-1} (\maxk + 1).
\end{eqnarray} \normalsize
The first equality is the rising sum of binomial coefficients and the rest is simple calculus.

\section{From AME to PA} \label{AMEToPAProof}
\textbf{Proof}: We consider the general case of the AME equation in \eqref{ame} for $|S|$ states and $N_I$ + $N_C$ rules. Assuming that the number of links of an $(s,k)$ node to all other states is multinomially distributed, we get two equations: The first equation describes the fraction of $(s,k)$ nodes and the second equation the probability $p_{s,k}[s_n]$ that a random neighbor of $(s, k)$ is in state $s_n$.

	\( \circ \) In order to derive an expression for the fraction of $(s,k)$ nodes,  we sum up both parts of \eqref{ame} over all possible vectors $\vecm \in \Vecm_k, $ where $\Vecm_k$ is the support of the multinomial distribution, and we apply the multinomial assumption to all terms. 
	This yields \small
	\begin{eqnarray}
	\label{ameToPA1}
	&&\frac{\partial}{\partial t}  \sumOverM x_{s, k, \vecm} = \nonumber \\
	&&\sum_{R_{I,i} \in R^{s^{+}}_{I}} \lambda_i  \sumOverM x_{r_i, k,\vecm} -\sum_{R_{I,i} \in R^{s^{-}}_{I}} \lambda_i \sumOverM  x_{s, k, \vecm} \nonumber\\
	&+& \sum_{R_{C,j} \in R^{s^{+}}_{C}} \lambda_j  \sumOverM x_{r_{j_1}, k, \vecm} \vecm[r_{j_2}] \nonumber\\
	&-&\sum_{R_{C,j} \in R^{s^{-}}_{C}} \lambda_j \sumOverM  x_{s, k, \vecm}  \vecm[r_{j_{2}}] \\ 
	&+&\sum_{\substack{ (s',s'') \in S^2\\ s'\neq s''}}  \betaS \!\!\!\!\!\!\sum_{\substack{ \vecm \in \Vecm_k\\ \vecm[s''] \geq 1, \vecm[s'] <k }} \!\!\!\!\! x_{s, k,\vecmS} \vecmS[s'] \nonumber\\
	&-& \sum_{\substack{ (s',s'') \in S^2 \\ s'\neq s''}} \betaS \!\!\!\!\sumOverM x_{s, k,\vecm}  \vecm[s'], \hspace{0.5cm} \forall s\in S, \;\forall k \in K, \nonumber
	\end{eqnarray} \normalsize
	which after simplification becomes 
	\small
	\begin{eqnarray}
	\label{ameToPA}
	&&\frac{\partial}{\partial t}   x_{s, k} = \sum_{R_{I,i} \in R^{s^{+}}_{I}} \lambda_i   x_{r_i, k} -\sum_{R_{I,i} \in R^{s^{-}}_{I}} \lambda_i  x_{s, k}\\
	&+& \sum_{R_{C,j} \in R^{s^{+}}_{C}} \lambda_j x_{r_{j_1},k }\firstMoment{r_{j_1},k }{\rjtwo} \quad -\sum_{R_{C,j} \in R^{s^{-}}_{C}} \lambda_j x_{s,k} \firstMoment{s,k}{\rjtwo}. \nonumber
	\end{eqnarray} \normalsize
	Here, $\firstMoment{s,k}{s_i}$ denotes the average of the
	$i$-th entry over all $\vecm$ vectors of $(s,k)$ nodes. Applying the multinomial assumption, it holds that $\firstMoment{s,k}{s_i} = \sumMk f(\vecm, p_{s,k}) \vecm[s_i] = k p_{s,k}[s_i],$ where $f(\vecm, p_{s,k})$ is the probability of picking a neighbor vector $\vecm$ from a multinomial distribution with a vector of success probabilities equal to $p_{s,k}.$
	
	Note that the last two terms  in \eqref{ameToPA1} cancel out because  \small
	\begin{eqnarray}
	&&\sum_{\substack{ \vecm \in \Vecm_k \\ \vecm[s'] <k \\\vecm[s''] \geq 1 }} \; x_{s, k,\vecmS} (\vecm[s'] +1) = \sum_{\substack{ \vecm \in \Vecm_k \\ \vecm[s'] \geq 1 \\\vecm[s''] <k }} \; x_{s, k,\vecm} \vecm[s'] = \nonumber\\
	&&\sum_{\substack{ \vecm \in \Vecm_k }} \; x_{s, k,\vecm} \vecm[s'],
	\end{eqnarray} \normalsize
	where the last equality   follows from the
	fact that when $\vecm[s''] =k$ then $\vecm[s'] =0$. 
	
	 $\circ $ To derive an equation for   $p_{s,k}[s_n]$, we first consider the change over time  of the
	number of edges between a node in state $s$ and $s_n$.  From the multinomial assumption, we have that \small
	\begin{eqnarray}
	\label{ameToPA21}
	\frac{\partial}{\partial t}  \sumOverM x_{s,k,\vecm} \vecm[s_n] &=& \frac{\partial}{\partial t} \bigg( x_{s,k} \sumOverM f(\vecm, p_{s,k}) \vecm[s_n]\bigg ) \nonumber \\
	&=& \partialt (x_{s,k} k p_{s,k}[s_n]),
	\end{eqnarray} \normalsize
	where $f(\vecm, p_{s,k})$ is the probability of picking a neighbor vector $\vecm$ from a multinomial distribution with a vector of success probabilities equal to $p_{s,k}.$ Using similar arguments, we can now derive an expression for \eqref{ameToPA21} by summation of     \eqref{ame} after multiplication of  $\vecm[s_n]$.   
	\small
	\begin{eqnarray}
	\label{ameToPA22}
	&&\qquad \partialt (x_{s,k} k p_{s,k}[s_n]) =\nonumber\\ 
	&& \sum_{R_{I,i} \in R^{s^{+}}_{I}} \lambda_i  \sumOverM x_{r_i, k} f(\vecm, p_{r_i,k} ) \vecm[s_n] \nonumber \\
	&&-\sum_{R_{I,i} \in R^{s^{-}}_{I}} \lambda_i \sumOverM  x_{s, k} f(\vecm, p_{s,k} )\vecm[s_n] \nonumber \\
	&&+ \sum_{R_{C,j} \in R^{s^{+}}_{C}} \lambda_j  \sumOverM x_{r_{j_1}, k} f(\vecm, p_{\rjonek} ) \vecm[s_n] \vecm[r_{j_{2}}] \nonumber\\ 
	&&-\sum_{R_{C,j} \in R^{s^{-}}_{C}} \lambda_j \sumOverM  x_{s, k} f(\vecm, p_{s,k} ) \vecm[s_n] \vecm[r_{j_{2}}] \nonumber\\ 
	&&+\!\!\!\!\!\sum_{\substack{ (s',s'') \in S^2\\ s'\neq s''}} \!\!\!\!\! \betaS \Bigg( \!\!\underbrace{\sum_{\substack{ \vecm \in \Vecm_k \\ \vecm[s'] <k\\ \vecm[s''] \geq 1 }} \!\!\! x_{s, k, \vecmS}  \vecmS[s'] \vecm[s_n]}_{(1)}  \nonumber \\
	&&-\underbrace{\sumOverM x_{s, k, \vecm}  \vecm[s'] \vecm[s_n]}_{(2)} \Bigg )
	\end{eqnarray} \normalsize
	For the last two sums, we distinguish the cases (i) $s_n \neq s' \neq s''$ and (ii) $s_n = s' \vee s_n = s''.$	

	In case (i) we get \small 
	\begin{eqnarray}
	 &&\sum_{\substack{ \vecm \in \Vecm_k\\ \vecm[s'] <k \\ \vecm[s''] \geq 1 }}   x_{s, k,\vecmS}  \big(\vecm[s']+1 \big) \vecm[s_n]  = \\
	 &&\sum_{\substack{ \vecm \in \Vecm_k\\ \vecm[s'] \geq 1 \\ \vecm[s''] <k }}  x_{s, k, \vecm}  \vecm[s'] \vecm[s_n],
	 \end{eqnarray}
	  \normalsize which implies that the difference (1)-(2) is zero for this case.
	Note that the entry of $s_n$ in vector $\vecm$ remains the same during substitution, since $s_n \neq s' \neq s''$. 
	
	In case (ii) we distinguish again between the two subcases: (a) $s_n = s'$, and (b) $s_n = s''.$
	\\\\
	\noindent \underline{Subcase (a)}: The difference (1)-(2) becomes \small
	\begin{eqnarray}
	\label{ameToPA23}
	&&\textstyle\sum\limits_{\substack{ \vecm \in \Vecm_k\\ \vecm[s_n] <k \\ \vecm[s''] \geq 1 }} \!\!\!\! x_{s, k, \vecmSn}  (\vecm[s_n]\!+\!1) \vecm[s_n] - \!\!\!\sum\limits_{\substack{ \vecm \in \Vecm_k\\ \vecm[s_n] \geq 1 \\ \vecm[s''] \geq 0 }} \!\!\!\!x_{s, k, \vecm}  \vecm[s_n]^2  \nonumber \\
	&&\textstyle=\sum\limits_{\substack{ \vecm \in \Vecm_k\\ \vecm[s_n] \geq 1 \\ \vecm[s''] <k }} \!\!\!x_{s, k, \vecm}  \vecm[s_n] (\vecm[s_n] -1) - \sum\limits_{\substack{ \vecm \in \Vecm_k\\ \vecm[s_n] \geq 1 \\ \vecm[s''] \geq 0 }} x_{s, k, \vecm}  \vecm[s_n]^2  \nonumber\\
	&&\textstyle= - \sum\limits_{\substack{ \vecm \in \Vecm_k\\ \vecm[s_n] \geq 1 \\ \vecm[s''] \geq 0 }} x_{s, k, \vecm}  \vecm[s_n] = - \sum\limits_{\substack{ \vecm \in \Vecm_k\\ \vecm[s_n] \geq 1 \\ \vecm[s''] \geq 0 }} x_{s, k} f(\vecm, p_{s,k}) \vecm[s_n] \nonumber\\ 
	&&\textstyle = - x_{s,k} k p_{s,k}[s_n].
	\end{eqnarray} \normalsize
	\\[1ex]
	\underline{Subcase (b)}: Similar to subcase (a), the difference (1)-(2) becomes \small
	\begin{eqnarray}
	\label{ameToPA24}
	&&\sum_{\substack{ \vecm \in \Vecm_k\\ \vecm[s'] <k \\ \vecm[s_n] \geq 1 }} \!\!\! \!x_{s, k, \vecmS}  (\vecm[s']\!+\!1) \vecm[s_n] -\!\!\!\!\! \sum_{\substack{ \vecm \in \Vecm_k\\ \vecm[s_n] \geq 1 \\ \vecm[s_n] \geq 0 }} \!\!\!\! x_{s, k, \vecm}  \vecm[s'] \vecm[s_n] \!= \nonumber\\
	&&\sum_{\substack{ \vecm \in \Vecm_k\\ \vecm[s'] \geq 1 \\ \vecm[s_n] < k }} x_{s, k, \vecm}  \vecm[s']  (\vecm[s_n]+1) - \sum_{\substack{ \vecm \in \Vecm\\ \vecm[s_n] \geq 1 \\ \vecm[s_n] \geq 0 }} x_{s, k, \vecm}  \vecm[s'] \vecm[s_n] = \nonumber\\
	&&\sum_{\substack{ \vecm \in \Vecm\\ \vecm[s'] \geq 1 \\ \vecm[s_n] \geq 0 }}  x_{s,k,\vecm} \vecm[s'] = \sum_{\substack{ \vecm \in \Vecm\\ \vecm[s'] \geq 1 \\ \vecm[s_n] \geq 0 }} x_{s,k} f(\vecm, p_{s,k}) \vecm[s'] = \nonumber\\
	 &&x_{s,k} k  p_{s,k}[s'] 
	\end{eqnarray} \normalsize
	Thus, overall, dividing both parts of \eqref{ameToPA22} by $x_{s,k} k$, rearranging the term and doing all the possible algebraic simplifications, we get 
	\small  
	\begin{eqnarray}
	\label{ameToPA25}
	&&\partialt p_{s,k}[s_n]=  -\partialt x_{s,k} \frac{p_{s,k}[s_n]}{x_{s,k}} \nonumber\\
	&&+ \indSumPlus \frac{\lambda_i x_{r_i,k}}{x_{s,k} k} \sumOverM f(\vecm, p_{r_i,k}) \vecm[s_n] \nonumber\\
	&&- \indSumMinus \frac{\lambda_i}{ k } \sumOverM f(\vecm, p_{s,k}) \vecm[s_n] \\
	&&+ \contSumPlus \frac{\lambda_j x_\rjonek} {x_{s,k} k} \sumOverM f(\vecm, p_{\rjonek}) \vecm[\rjtwo] \vecm[s_n] \nonumber\\
	&&- \contSumMinus \frac{\lambda_j }{ k} \sumOverM f(\vecm, p_{s,k}) \vecm[\rjtwo] \vecm[s_n] \nonumber\\
	&&+ \sum_{\substack{ s' \in S\\ s'\neq s_n}} \betaSn p_{s,k}[s'] - \sum_{\substack{ s'\in S\\ s'\neq s_n}} \betaSnn p_{s,k}[s_n].\nonumber
	\end{eqnarray} \normalsize 
	Next we replace and get:   
	\small
	\begin{eqnarray}
	\label{ameToPA25}
	&&\partialt p_{s,k}[s_n] = -\partialt x_{s,k} \frac{p_{s,k}[s_n]}{x_{s,k}} \nonumber\\
	&&+ \indSumPlus \frac{\lambda_i x_{r_i,k}}{x_{s,k} k} \firstMoment{r_{i},k}{s_n} - \!\!\!\indSumMinus \frac{\lambda_i}{ k } \firstMoment{s,k}{s_n} \nonumber\\
	&&+ \contSumPlus \frac{\lambda_j x_\rjonek} {x_{s,k} k} \secMoment{\rjonek}{\rjtwo}{s_n}   \\
	&& - \contSumMinus \frac{\lambda_j }{ k} \secMoment{s,k} {\rjtwo }{s_n}  \nonumber\\
	&&+ \sum_{\substack{ s' \in S\\ s'\neq s_n}}  \betaSn p_{s,k}[s'] - \sum_{\substack{ s' \in S\\ s'\neq s_n}}  \betaSnn p_{s,k}[s_n]\nonumber .
	\end{eqnarray}     \normalsize
	Finally, after imposing the multinomial assumption in the computation of the rates $\betaS$,
	we get \small
	\begin{eqnarray}
	\label{ameToPABetas}
	&&\betaS = \nonumber\\
	&& \frac{\avg{\sumMk \vecm[s] x_{s',k,m} \big(\sum_{R^{s'\rightarrow s''}_{I}} \lambda_i  + \sum_{R^{s'\rightarrow s''}_{C}} \lambda_j \vecm[\rjtwo] \big) }} {\avg{\sumMk \vecm[s] x_{s',k,m}}} \nonumber\\
	&&= \frac{\avg{\sumMk \vecm[s] x_{s',k} f(\vecm, p_{s',k}) \big(\sum_{R^{s'\rightarrow s''}_{I} }\lambda_i  + \sum_{R^{s'\rightarrow s''}_{C}} \lambda_j \vecm[\rjtwo] \big) }} {\avg{\sumMk \vecm[s] x_{s',k} f(\vecm, p_{s',k}) }} \nonumber\\
	&&= \frac{\bigAvg{x_{s',k} \sum_{R^{s'\rightarrow s''}_{I}} \lambda_i  \firstMoment{s',k}{s}}  + \bigAvg{x_{s',k}  \sum_{R^{s'\rightarrow s''}_{C}} \lambda_j \secMoment{s',k}{s}{\rjtwo} }} {\avg{x_{s',k} \firstMoment{s',k}{s}}}. \nonumber
	\end{eqnarray}    \normalsize

For $s, s_n, s'_n\in S$ and $k \in K$, we have the following:
\begin{itemize}[leftmargin=*]  
	\item The expectation of the multinomial distribution is
	$$\firstMoment{s,k}{s_n} = k p_{s,k}[s_n].$$
	\item For the second mixed moment   we distinguish the     cases
	\begin{enumerate}[label=(\roman*)] 
		\item $s_n \neq s'_n$
		\small
		\begin{align}\label{secMomentA}
		\begin{split}
		\secMoment{s,k}{s_n}{s_n'} &= \secCentralMoment{s,k}{s_n}{s_n'} + \firstMoment{s,k}{s_n} \firstMoment{s,k}{s_n'} \\
		&= -k p_{s,k}[s_n] p_{s,k}[s'_n] + k p_{s,k}[s_n] k p_{s,k}[s'_n]\\
		&= k(k-1) p_{s,k}[s_n]  p_{s,k}[s'_n]
		\end{split}
		\end{align} \normalsize     
		\item $s_n = s'_n$ \small
		\begin{align}\label{secMomentB}
		\begin{split}
		\secMoment{s,k}{s_n}{s_n} &=   \secCentralMoment{s,k}{s_n}{s_n} + \firstMoment{s,k}{s_n}^2 \\
		&= k p_{s,k}[s_n] (1 - p_{s,k}[s_n]) + k p_{s,k}[s_n] k p_{s,k}[s_n]\\
		&= k p_{s,k}[s_n] (1 - p_{s,k}[s_n]  + k p_{s,k}[s_n]), 
		\end{split}
		\end{align}     \normalsize
	\end{enumerate}
\end{itemize}
In the above equations $\secCentralMoment{s,k}{s_n}{s_n'}$ is the second central mixed moment, i.e., the covariance of the multinomial distribution for the number of edges between an $(s,k)$ node and  $s_n$ nodes and
between an $(s,k)$ node and  $s_n'$ nodes, while 
$\secCentralMoment{s,k}{s_n}{s_n}$ is the second central moment, i.e., the variance for the number of edges between an $(s,k)$ node and  $s_n$ nodes.

\section{Simplification of sums for lumping PA}\label{sumsInPALumping}
To simplify the sum $\sumKBin \frac{1}{k} \secMoment{s,k}{r_{j_2}}{s_n}  P(k|b)$ of Eq.\,\eqref{PAProbLumpedFinal}, we consider the two cases of \eqref{secMomentA} and \eqref{secMomentB} for the second mixed moment and we get:
\begin{enumerate}[label=(\roman*)] 
	\item $s_n \neq s'_n$
	\small
	\begin{eqnarray}\label{sumA}
	&&\sumKBin \frac{1}{k} \secMoment{s,k}{s_n}{s'_n}  P(k|b) = \nonumber \\
	&& \sumKBin \frac{1}{k} k(k-1) p_{s,k}[s_n]  p_{s,k}[s'_n]  P(k|b)  =\nonumber\\
	&&\sumKBin (k-1) p_{s,b}[s_n]  p_{s,b}[s'_n] P(k|b)   = \nonumber\\
	&& p_{s,b}[s_n]  p_{s,b}[s'_n] \avg{k-1}_b 
	\end{eqnarray} \normalsize    
	\item $s_n = s'_n$ \small
	\begin{eqnarray}\label{sumB}
	&&\sumKBin \frac{1}{k} \secMoment{s,k}{s_n}{s_n}  P(k|b) = \nonumber \\
	&& \sumKBin \frac{1}{k}  k p_{s,k}[s_n] (1 - p_{s,k}[s_n]  + k p_{s,k}[s_n])  P(k|b)  =\nonumber\\
	&&\sumKBin p_{s,k}[s_n] P(k|b) + \sumKBin  p_{s,k}[s_n]^2(k-1)P(k|b) = \nonumber\\
	&& p_{s,b}[s_n] + p_{s,b}[s_n] ^2 \avg{k-1}_b
	\end{eqnarray}
   \normalsize
\end{enumerate}
Similarly, for simplifying $\secMoment{s,b}{s_n}{s_n'}$ of Eq.\,\eqref{eq:assumimpl2} that is used
for the computation of $\betaS_{B}$, we get:
\begin{enumerate}[label=(\roman*)] 
	\item $s_n \neq s'_n$
	\small
	\begin{eqnarray}\label{sumA}
	&&\sumKBin  \secMoment{s,k}{s_n}{s'_n}  P(k|b) = \nonumber \\
	&& \sumKBin  k(k-1) p_{s,k}[s_n]  p_{s,k}[s'_n]  P(k|b)  =\nonumber\\
	&& p_{s,b}[s_n]  p_{s,b}[s'_n] \avg{k(k-1)}_b 
	\end{eqnarray} \normalsize    
	\item $s_n = s'_n$ \small
	\begin{eqnarray}\label{sumB}
	&&\sumKBin  \secMoment{s,k}{s_n}{s_n}  P(k|b) = \nonumber \\
	&& \sumKBin   k p_{s,k}[s_n] (1 - p_{s,k}[s_n]  + k p_{s,k}[s_n])  P(k|b)  =\nonumber\\
	&&\sumKBin k p_{s,k}[s_n] P(k|b) + \sumKBin  p_{s,k}[s_n]^2 k(k-1)P(k|b) = \nonumber\\
	&& \avgk_b p_{s,b}[s_n] + p_{s,b}[s_n] ^2 \avg{k(k-1)}_b
	\end{eqnarray}
	\normalsize
\end{enumerate}

\section{DBMF dynamics for SIR model} \label{dynamics}
In Fig.\,\ref{fig:sir_extraDynamics} we show the dynamics of the SIR models in several networks with $\maxk \in \{10^2, 10^3, 10^5, 10^6\}$. The degree distribution of each of the networks was chosen such that the coefficient of variation $\sigma / \avgk$ of the distribution remains constant. For a truncated power law probability distribution $P(k) = ck^{-\alpha}, k \in [\mink, \maxk]$ where $c$ is the normalizing constant $c=\frac{\alpha-1}{\mink^{-\alpha+1}},$ it holds that $\avgk = \alpha \Big[ \frac{k^{2-\alpha}}{2-\alpha} \Big]^{\maxk}_{\mink}$ and $\secMomK = \alpha \Big[ \frac{k^{3-\alpha}}{3-\alpha} \Big]^{\maxk}_{\mink}$. 
We get $\sigma = \sqrt{\secMomK - \avgk{^2}},$ and fixing $\sigma / \avgk = 7.5$ and $\mink = 1$, we compute $\alpha$ for each different $\maxk.$

\begin{figure}[h!]
	\centering
	\subfigure[]{
		\includegraphics[width=0.45\linewidth]{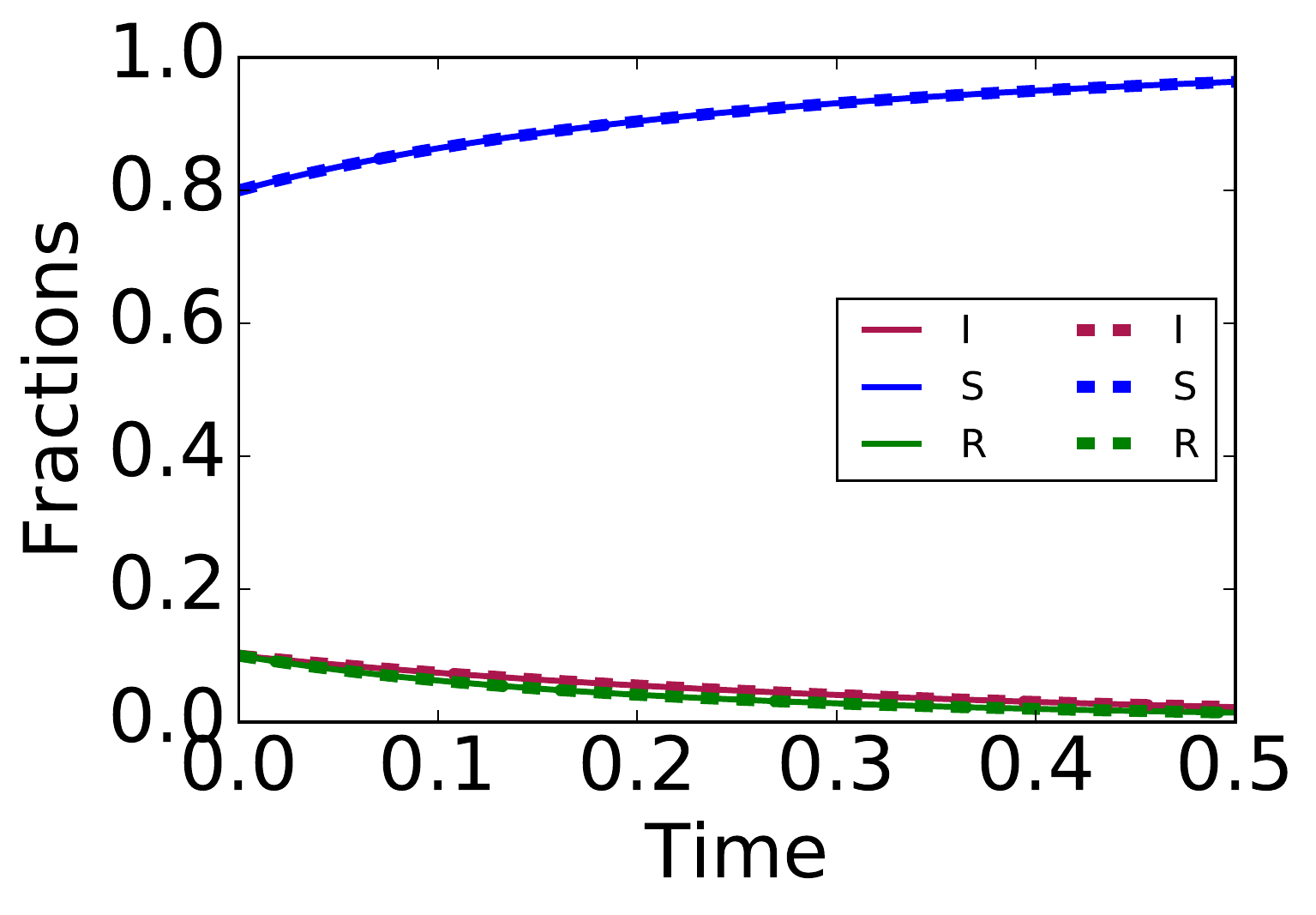} 
		\label{figapp:sir_dbmf_traj}
	}
	\subfigure[]{
		\includegraphics[width=0.45\linewidth]{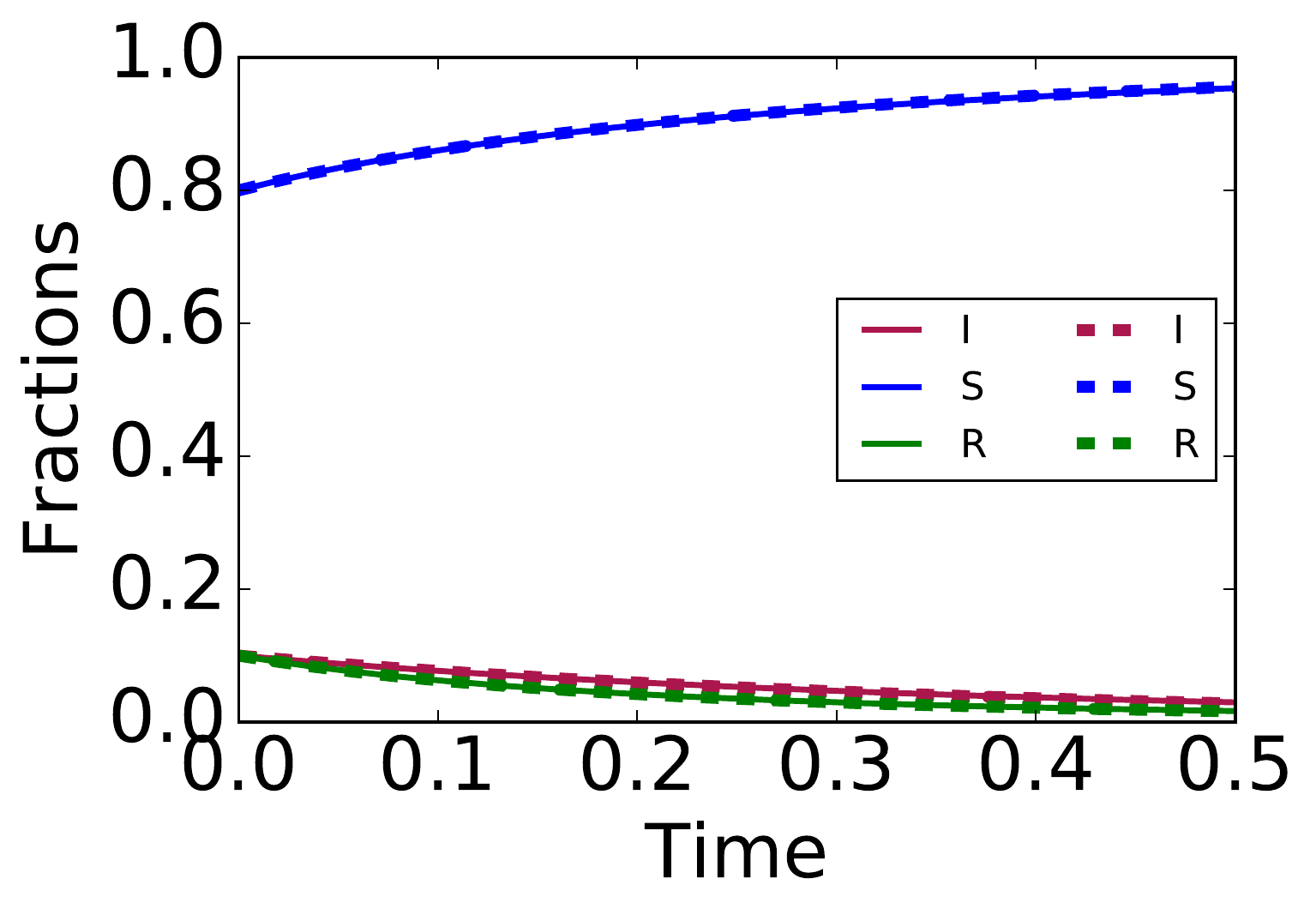} 
		\label{figapp:sir_dbmf_errors}
	}
	\subfigure[]{ 
		\includegraphics[width=0.45\linewidth]{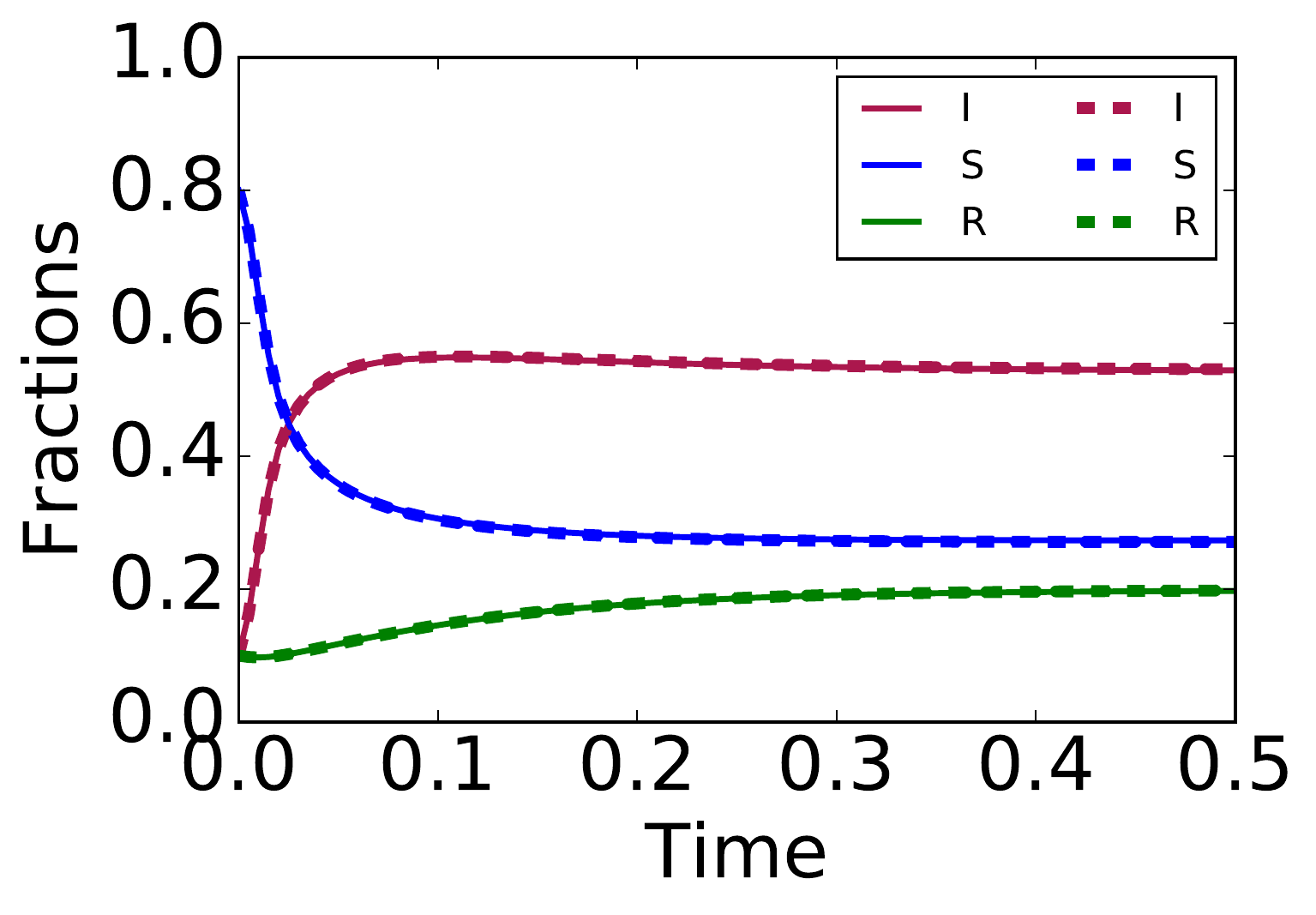} 
		\label{figapp:sir_pa_traj}
	}
	\subfigure[]{ 
		\includegraphics[width=0.45\linewidth]{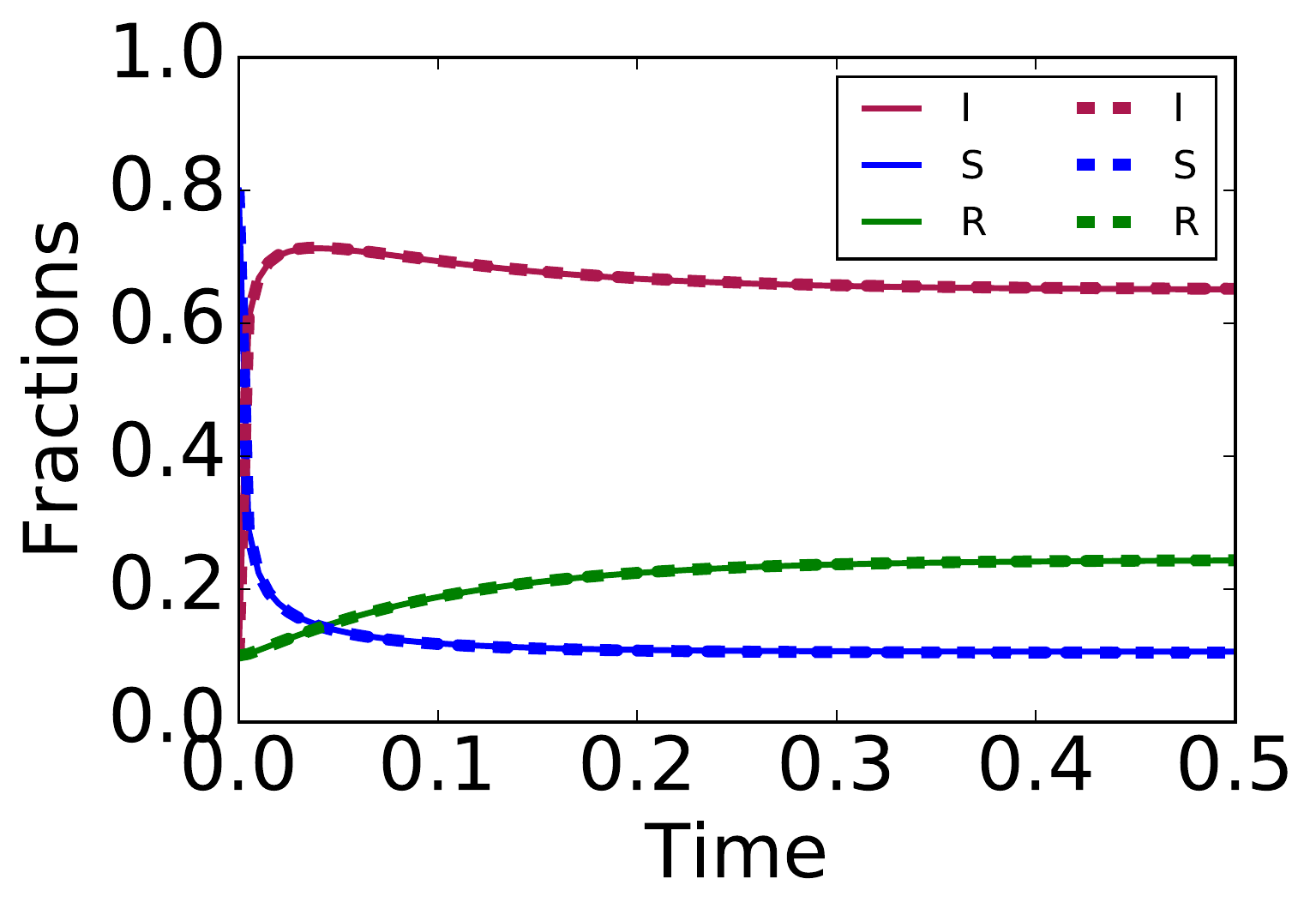} 
		\label{figapp:sir_pa_errors}
	}
	\caption{Total fraction of infected $(I)$, susceptible $(S)$ and recovered $(R)$ nodes of the SIR model in a power law network with (a): $\maxk = 10^2, P(k) \propto k^{-2.5},$ (b): $\maxk = 10^3, P(k) \propto k^{-1.14},$ (c): $\maxk = 10^5, P(k) \propto k^{-0.75},$ (d): $\maxk = 10^6, P(k) \propto k^{-0.69}$. The dense lines show the solution of the original DBMF equations and the dashed lines the solution of the lumped DBMF equations.}
	\label{fig:sir_extraDynamics}
\end{figure}

\section{Initial conditions for SIIIR model} \label{initialCond}
In Fig.\,\ref{fig:sir_extraDynamics} we show the PA lumping error of the SIIIR model in a power law network with $\maxk = 100, P(k) \propto k^{-2.5}$ over time for different choices of the initial fractions $x_{s,k},$ where $s\in \{S, I, II, III, R\}$ and $k \in K$. We plot the error of the initial conditions we used for the results of Fig.\,\ref{fig:siiir_pa} (our initial conditions), the error of an initialization that gives the same values to all $x_{s,k}$ for all $s$ and $k$ (common uniform), and finally we plot the average error of 30 different random initial conditions for the fractions $x_{s,k}$ together with the error's sample standard deviation $(\pm \sigma).$ 
\begin{figure}[h!]
	\centering
		\includegraphics[width=0.6\linewidth]{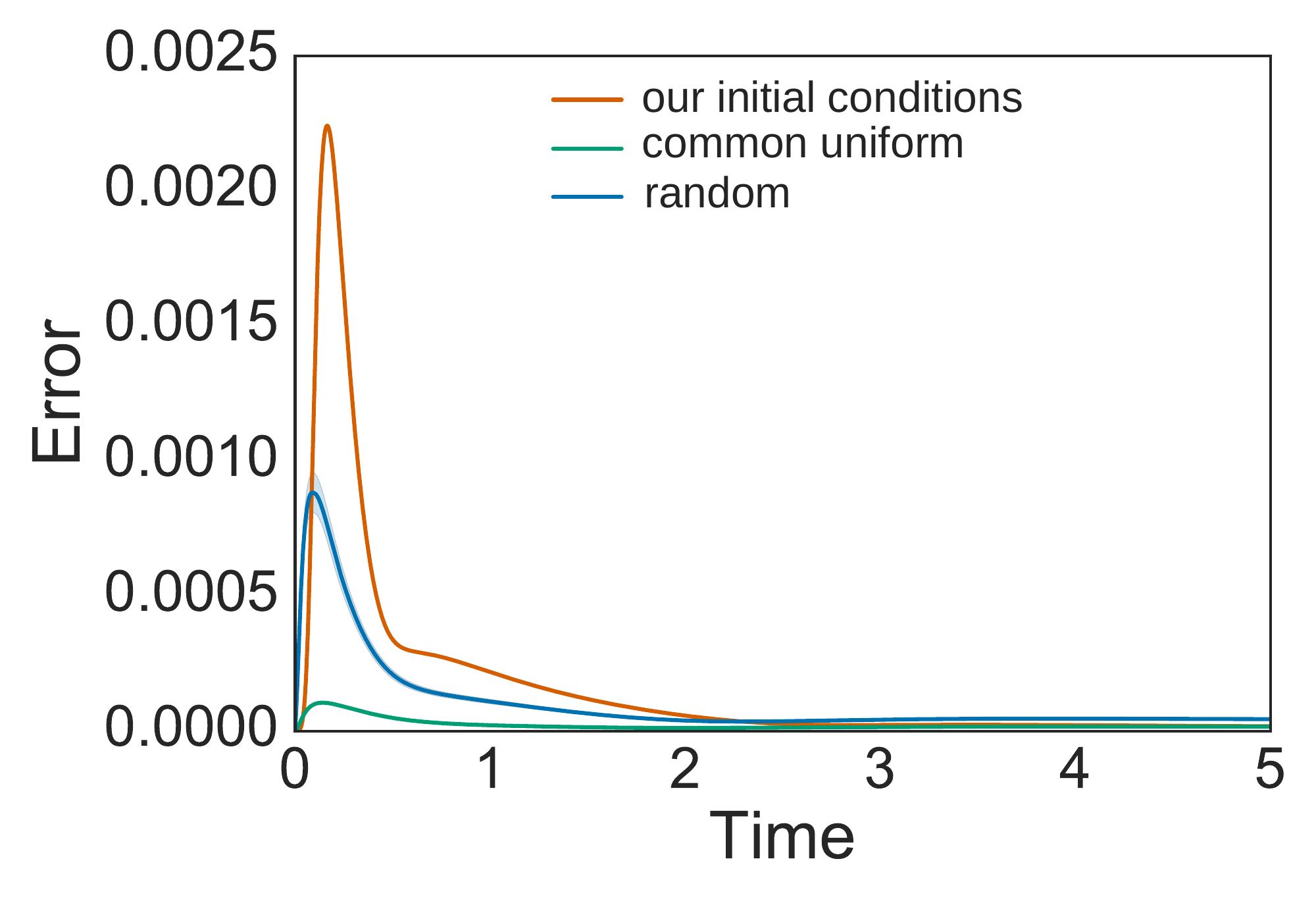} 
		\label{figapp:initialCond}

	\caption{The error $\epsilon(n)$ for $n=17$ for the PA lumped equations considering different initial conditions.}  
	\label{fig:sir_extraDynamics}
\end{figure}

\bibliographystyle{apsrev4-1}
\bibliography{../sample}

\end{document}